\documentstyle[epsfig]{mn2e}

\begin{document}

\title[Mass modelling of Superthin Galaxies]
{Mass modelling of Superthin Galaxies: IC5249, UGC7321, IC2233}
\author[Banerjee \& Bapat]
       {Arunima Banerjee$^{1}$\thanks{E-mail : arunima@iucaa.in} and 
        Disha Bapat$^{2}$\thanks{E-mail : dishabapat@gmail.com} \\
$^1$  The Inter-University Centre for Astronomy and Astrophysics, Pune 411007, India \\ 
$^2$ Department of Physics, Fergusson College, University of Pune 411007, India \\} 
\maketitle

\begin{abstract}

\noindent Superthin galaxies are low surface brightness disc galaxies, characterised by optical discs with strikingly high values of 
planar-to-vertical axes 
ratios ($>$ 10), the physical origin and evolution of which continue to be a puzzle.
We present mass models for three superthin galaxies: IC5249, UGC7321 and IC2233. 
 We use high resolution rotation curves and gas surface density distributions obtained from HI 21 cm radio-synthesis observations, in 
combination with their two-dimensional structural surface brightness decompositions at Spitzer 3.6 $\mu$m band, all of which were available
 in the literature. We find that while models with the pseudo-isothermal (PIS) and the Navarro-Frenk-White (NFW) dark matter density 
profiles fit the observed rotation curves of IC5249 and UGC7321 equally well, those with the NFW profile does not comply with the 
slowly-rising rotation curve of IC2233. Interestingly, for all of our sample galaxies, the best-fitting mass models with a PIS dark matter 
density profile indicate a {\it compact} dark matter halo i.e., $R_c/R_D$ $<$ 2 where $R_c$ is the core radius of the PIS dark matter halo, and
 $R_D$ is the radial scale-length of the exponential stellar disc. The compact dark matter halo may be fundamentally responsible for the 
superthin nature of the stellar disc, and therefore our results may have important implications for the 
formation and evolution models of superthin galaxies in the universe.

\end{abstract}

\section{Introduction}

\noindent Superthin galaxies are low surface brightness edge-on galaxies
with ''razor-thin" optical discs characterised by unusually high values of vertical-to-planar axes ratio ($>$ 10) with little or 
no discernable bulge component. They are typically gas-rich and dark matter dominated 
with very low star formation rates, low metallicity and dust content, and are classic examples of  
under-evolved, late-type galaxies (See Matthews, Gallagher \& van Driel 1999; Kautsch 2009 for a review).  
The term ''superthin" was coined by Goad \& Roberts (1981) who did spectroscopic studies of a group of four such galaxies: 
UGC 4278, 7170, 7321, and 9242. Karachentsev, Karachentseva \& Parnovskij (1993) conducted a systematic search for disc-like edge-on-galaxies 
with diameter a $>$ 40 $\arcsec$ and a major-to-minor axis ratio a/b $>$ 7 using the Palomar Observatory Sky Survey and the ESO/SERC 
survey, which culminated in the Flat Galaxy Catalogue (FGC) containing 4,455 flat or bulge-less galaxies covering 56$\%$ of the whole sky, and 
eventually the Revised Flat Galaxy Catalogue (RFGC) containing 4,236 galaxies covering the whole sky (Karachentsev et al. 1999). Superthins 
 constitute a special category of the FGC galaxies with major-to-minor axes ratio a/b $>$ 10, which have also been extensively
studied later as part of other surveys related to edge-on, disc galaxies in general (Dalcanton \& Bernstein 2002; Kregel, van der Kruit \& Freeman 2005; Kautsch et al. 2006; Yoachim \& Dalcanton 2006; Comeron et al. 2011; Bizyaev et al. 2014).\\

\noindent Being rich in neutral hydrogen gas (HI) content, superthin galaxies have been observed in the HI 21 cm radio observations as well. 
Matthews \& van Driel (2000) conducted a survey to measure the HI content of a sample of 400 gas-rich FGC galaxies 
which confirmed rich reserves of HI in them. However, HI 21 cm radio-synthesis
 observations aimed at studying distribution and kinematics of HI in superthins have been sparse (But, see, van der Kruit et al. 2001; 
Uson \& Matthews 2003; 
O'Brien et al. 2010a). Since the size of the HI disc in spiral galaxies is in general a few times the size of the optical disc, HI may act as a 
useful diagnostic tracer of the underlying gravitational potential of the galactic dark matter halo. In fact, for edge-on or moderately edge-on 
galaxies, HI 21 cm spectroscopic observations may be used to determine the galactic rotation curve, which, in combination with stellar surface 
photometry studies, can be used to construct galaxy mass models, thereby determining the density profile of the dark matter halo.  
(de Blok et al. 2001, 2008; Oh et al. 2008, 2015). In addition, HI vertical scale-height data  may be used to 
constrain the vertical component of the underlying gravitational field, and thereby the flattening of the 
galactic dark matter halo density distributions (See, for example, Banerjee \& Jog 2011 and the references therein). 
According to the modern theories of galaxy 
 formation and evolution, galaxies form and evolve at the centres of the gravitational potentials of their dark matter haloes, 
which therefore play a crucial role in regulating the disc structure and dynamics. Using the joint constraints of the HI rotation curve and 
the HI scale-height data, Banerjee, Matthews \& Jog (2010) found that the superthin
galaxy UGC 7321 has a dense and compact dark matter halo  i.e., $R_c/R_D$ $<$ 2 where $R_c$ is the core radius of the dark matter halo, and
 $R_D$ is the radial scale-length of the stellar disc (See, also, O'Brien, Freeman \& van der Kruit. 2010d). Using their 2-component model of gravitationally-coupled
 stars and gas in the force field of a dark matter halo, Banerjee \& Jog (2013) further showed that the gravitational potential of the compact 
 dark matter halo is primarily responsible for the superthin vertical structure of the stellar disc in UGC7321. This was a vital 
 clue towards developing a comprehensive theory of the origin and evolution of superthin galaxies. However we do not know if a compact dark 
matter halo is a generic
 feature of superthin galaxies as detailed mass modelling studies of superthins have been lacking until the current work. \\

\noindent In this paper, we therefore construct detailed mass models for three superthin galaxies viz. IC5249, UGC7321 and IC2233 for which high 
resolution HI rotation curves, HI radial surface density profiles as well as structural surface brightness decompositions of Spitzer 
3.6 $\mu$m images were available in the literature, with the main aim of understanding the density profile of their dark matter haloes. 
The rest of the paper is organised as follows: In \S 2, we describe our sample, in \S 3 the data and the 
input parameters, in \S 4 we discuss the mass models, followed by the results and discussion \S 5 and finally the conclusions in \S 6.

\section{Sample}
\noindent We consider a sample of three galaxies for our mass modelling studies viz. IC5249, UGC7321 and IC2233, which are   
classic examples of superthin galaxies. Out of these, UGC7321 and IC2233 were two of the four superthins originally observed by 
Goad \& Roberts 1981 (\S 1). UGC7321, in particular, was studied in detail as a prototypical superthin, both in  
multi-wavelength observations including the optical, mid-infrared as well as in HI (Matthews, Gallagher \& van Driel. 1999, Matthews 2000, Uson \& Matthews 2003, 
Matthews \& Wood 2003). The shape and density profile of its dark matter halo was also modelled using the joint constraints of the HI rotation 
curve and the HI vertical scale-height (Banerjee et al. 2010, O'Brien et al. 2010d), which revealed a dense and compact pseudo-isothermal dark 
matter halo (\S 4.2.2). Among our other two sample galaxies, the kinematics of the stellar disc of IC5249, another prototype and also one of the 
flattest or thinnest galaxies ever known, was investigated using 
stellar spectroscopy and HI observations, which indicated a stellar disc as \emph{hot} as that of the Galaxy, contrary to the expectation of
 an ultra-cold stellar disc as implied by its razor-thin appearance (van der Kruit et al. 2001). Since the disc vertical thickness is also
 strongly regulated by the mass distribution of the different gravitating components in terms of the resulting vertical gravitational field, 
it is crucial to obtain the mass models of these galaxies to understand their origin and evolution. However, detailed 
mass modelling of these superthins have not been attempted so far (\S 1) although high resolution HI rotation curves were available in the 
literature. In this paper, we have obtained mass models of these three superthin galaxies, using their 
HI rotation curves and the HI surface density along with structural surface brightness decompositions of Spitzer 3.6 $\mu$m images, all of which
 were available in the literature. \\

\noindent Our sample constitutes nearby galaxies located within a distance D $\sim$ 10 - 30 Mpc. The galaxies are highly inclined or edge-on 
with inclination angle $i$ $\sim$ 88$^o$ - 90$^o$ and have high values of major-to-minor axes ratio a/b $\sim$ 7 - 10.3. Finally, they are
 low surface brightness with a de-projected B-band central surface brightness ${\mu}_{0,B}$ $\sim$ 22.6 - 24.5 mag arcsec$^{-2}$.
We summarise the basic properties of our sample galaxies in Table 1.

\begin{table}
\begin{center}
\begin{minipage}{150mm}
{\small
\hfill{}
\caption{Basic properties of the sample superthin galaxies}
\centering
\begin{tabular}{l|c|c|c|c}
\hline
Galaxy & Distance & Inclination & a/b \footnote{major-to-minor axes ratio} & ${{\mu}_{0,B}}$\footnote{Face-on central surface brightness in B-band}           \\

       & (Mpc)    & ($^o$)      &                 &      (mag arcsec$^{-2}$)                                \\
\hline

IC5249\footnote{Values quoted from van der Kruit et al. (2001)} & 30.4      &  90    & 10.2                          &       24.5           \\
UGC7321\footnote{Values quoted from O'Brien et al. (2010d)}& 10      &  88 $\pm$ 1     & 10.3           &             23.5                  \\
IC2233\footnote{Values quoted from Matthews \& Uson (2008)} & 10      &  88.5 $\pm$ 1.5 & 7          &             22.6                    \\

\hline
\end{tabular}}
\hfill{}
\label{tb:tablename}
\end{minipage}
\end{center}
\end{table}

\section{Data \& Input Parameters}

\subsection{Stellar Photometry}

\noindent We get the radial surface brightness profiles for our sample galaxies from the two-dimensional structural surface 
brightness decompositions at Spitzer 3.6 $\mu$m band for the complete sample of S$^4$G galaxies 
(Salo et al. 2015). Mid-infrared emission traces the old stellar 
population dominant in late-type galaxies, whereas optical photometry indicates the young stellar population which constitutes only 
a small fraction of the total stellar dynamical mass. Besides, the primary source of error in deriving stellar mass from observed
 surface brightness distribution lies in obtaining the stellar mass-to-light ratio (${{\Upsilon }_{\star}}$)  
from the population synthesis models (\S 4.3), which gets complicated by dust extinction issues predominant in edge-on galaxies. Unlike the
 optical, mid-infrared emission is less prone to dust extinction, and is also unaffected by recent star formation, young stellar population and 
 choice of the Initial Mass Function (IMF) in the stellar population synthesis model. Therefore, it is commonly preferred over optical photometry to model the galactic stellar disc in mass-modelling studies (de Blok et al. 2008; Oh et al. 2008; Oh et al. 2015). \\ 

\noindent  Salo et al. (2015) performed multi-component decompositions of the surface brightness profiles of the complete sample of S$^4$G 
galaxies using GALFIT 3.0. 
The surface brightness profiles of the edge-on and the nearly edge-on galaxies were fitted with the GALFIT function \emph{edgedisk}, 
which is that of an exponential disc with a finite vertical thickness as viewed edge-on, and therefore
has four free parameters: the face-on, central surface brightness (${\mu}_0$),  radial scale-length ($R_D$), vertical scale -length ($h_z$) and the 
position  angle  of  the disc. 
In some cases, a bulge, a bar or an additional thick disc component
were also included. In Figure 1, we present the inclination-corrected radial surface brightness profiles for our sample galaxies,
obtained from the multi-component decompositions of their 3.6 $\mu$m surface brightness profiles as given in the above paper. 
All three of our sample galaxies were found to be consistent with a model comprising superposition of 
 two edge-on discs: the \emph{outer disc} and the \emph{inner disc}. Henceforth, we will denote the structural parameters of the \emph{outer disc} 
 with a suffix 1 (${\mu}_0$(1),  $R_D$(1), $h_z$(1)), and those of the \emph{inner disc} with a suffix 2 (${\mu}_0$(2),  $R_D$(2), $h_z$(2)).
Interestingly, unlike in the optical, the stellar discs of our
sample galaxies are not low surface brightness in nature in mid-infrared. Also, the high values of 
the vertical-to-planar scale-length ratios $h_z(1)/R_D(1)$ $>$ 0.1 indicates that the stellar disc is also not superthin either in this band.
In general, the \emph{outer discs} of our sample galaxies have a relatively lower surface brightness
(${\mu}_0$(2) $>$ 21 mag arcsec$^{-2}$) with a larger radial scale-length ($R_D$(1) $\sim$ 2-5 kpc).
The \emph{inner discs}, on the other hand, have a high de-projected disc central surface brightness (${\mu}_0$(2) $<$ 20 mag 
arcsec$^{-2}$) with a smaller radial scale-length ($R_D$(2) $\sim$ 1 kpc). 
The \emph{outer disc} can therefore be taken to be the main stellar disc component of the galaxy.
Similarly, since the \emph{inner disc} constitutes only a small fraction of the total model flux, it is
 arguable if it can be taken to be the bulge or central spheroidal component of the host galaxy.
(See \S 5.2 de Blok et al. 2008; \S 3 of Salo et al. 2015, for discussion).
Finally, as the radial profile of the inner disc is steep and that of the outer one shallow, the stellar discs of 
these galaxies are considered to be anti-truncated. \\

\begin{figure*}
\begin{center}
\begin{tabular}{cccc}
\resizebox{40mm}{!}{\includegraphics{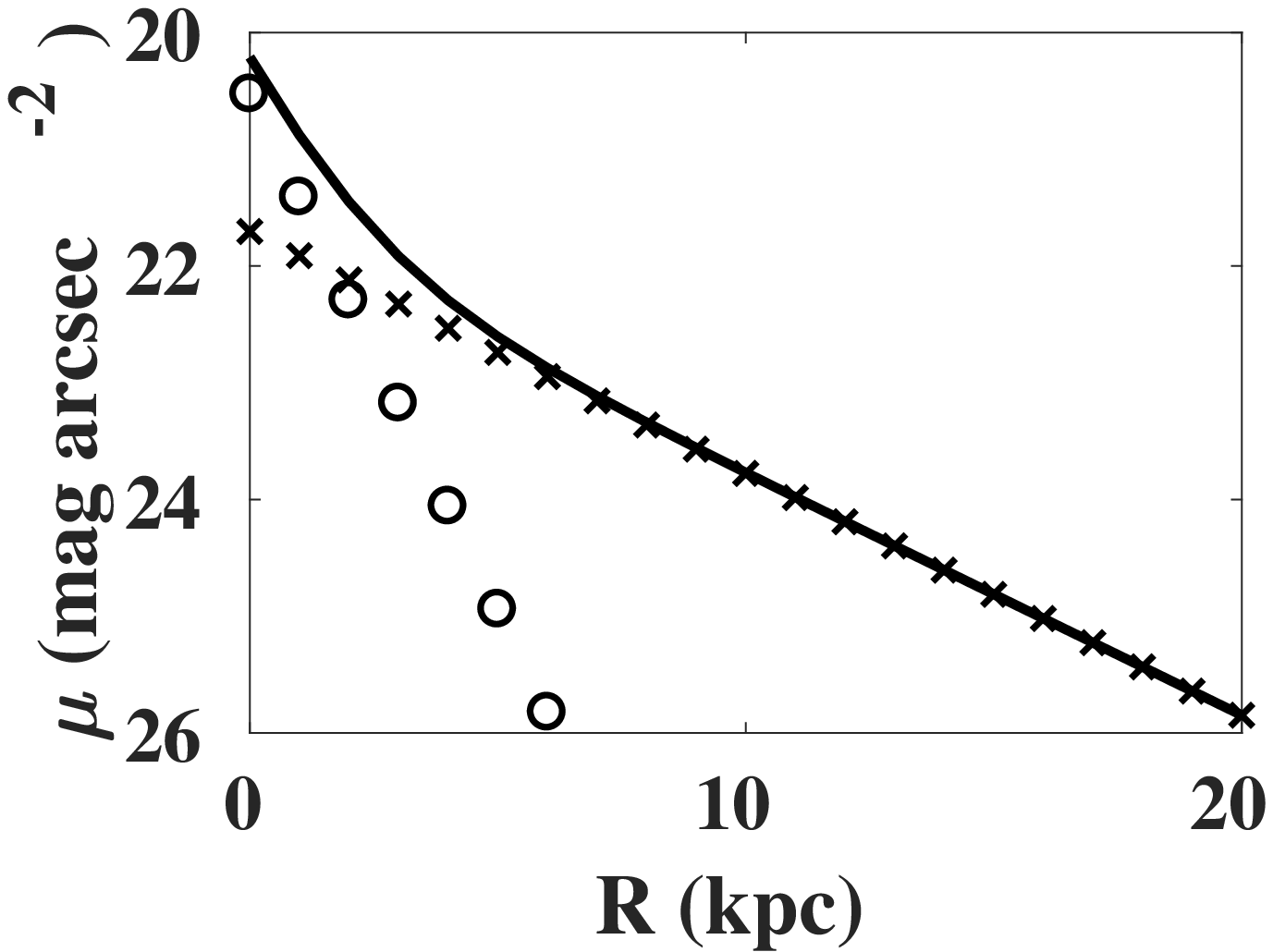}} &
\resizebox{40mm}{!}{\includegraphics{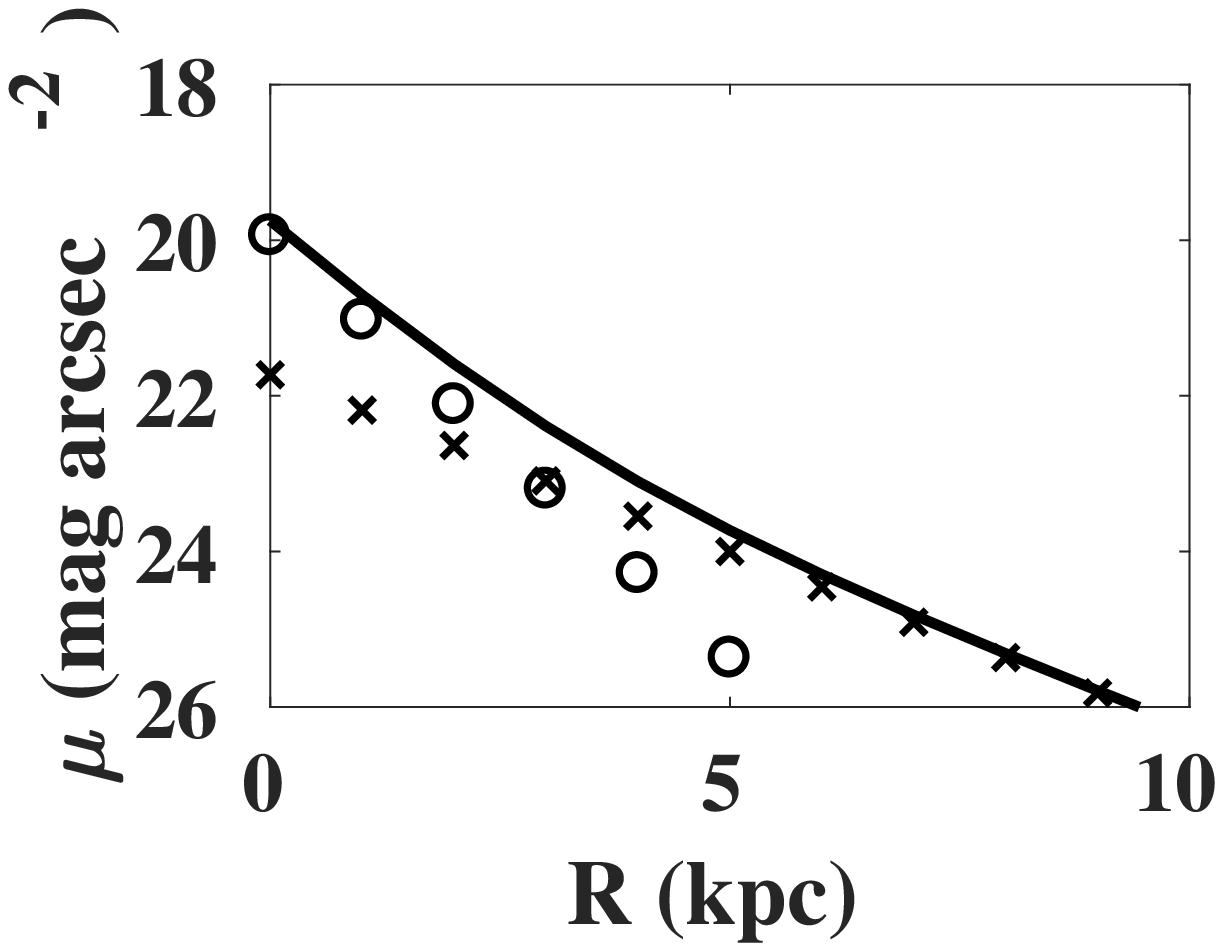}} &
\resizebox{40mm}{!}{\includegraphics{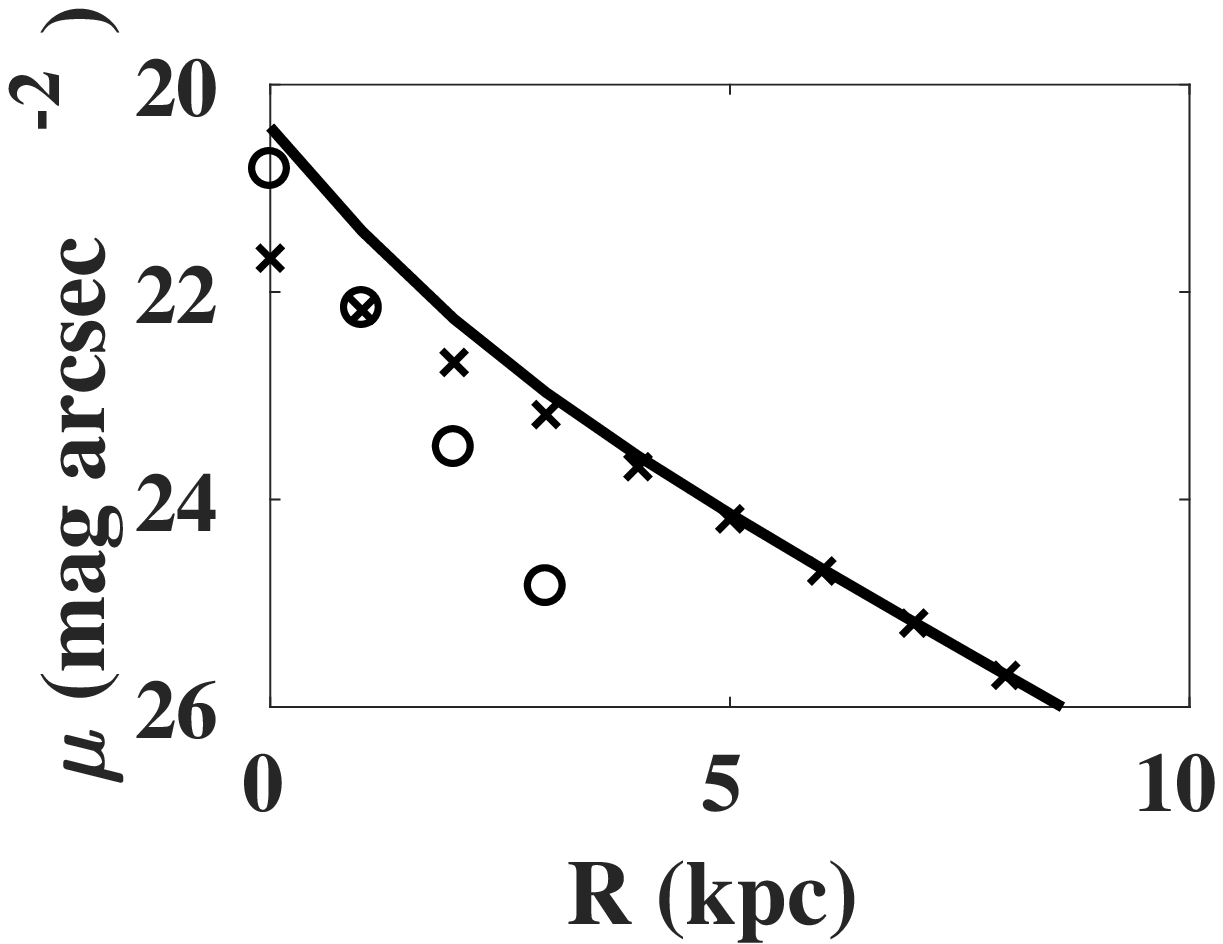}} \\
\end{tabular}
\end{center}
\caption{Radial surface brightness profiles for our sample galaxies as given by the two-dimensional structural surface brightness decompositions 
at Spitzer 3.6 $\mu$m emission (Salo et al. 2015). The solid line indicates the total surface brightness profile while the contributions of the 
 inner and the outer discs are given by the \emph{open circles} and the \emph{crosses} respectively: Panel [1] IC5249, Panel [2] UGC 7321 and Panel [3] IC2233}. \label{fig:mu}
\end{figure*}

\noindent Our sample galaxies have almost identical values of the central surface brightness (${\mu}_0$(2) $\sim$ 21 mag arcsec$^{-2}$) of their outer 
discs. As far as the \emph{inner discs} are concerned, UGC7321 has the highest central surface brightness 
(${\mu}_0$(1) $\sim$ 19 mag arcsec$^{-2}$), which is an order of magnitude higher than those of IC5249 and IC2233. 
(${\mu}_0$(2) $\sim$ 20 mag arcsec$^{-2}$). Finally, 
with regard to the size of the \emph{inner disc}, IC5249 has the smallest one ($R_D(2)/R_D(1)$ $\sim$ 0.23) while UGC7321 and IC2233 have 
relatively bigger \emph{inner disc} components ($R_D(2)/R_D(1)$ $\sim$ 0.37).  
We present the structural 
parameters of our sample galaxies at 3.6 $\mu$m surface brightness profiles in Table 2. \\

\begin{table}
\begin{center}
\begin{minipage}{125mm}
{\small
\hfill{}
\caption{Surface brightness decompositions at Spitzer 3.6$\mu$m band}
\centering
\begin{tabular}{l|c|c|c|c|c|c|c}
\hline
Galaxy \footnote{Values quoted from Salo et al. (2015)} & ${\mu}_0$(1) \footnote{Face-on surface brightness of outer disc in mag arcsec$^{-2}$} & $R_D$(1) \footnote{Radial scale-length of outer disc in kpc} & $h_z$(1)\footnote{Vertical scale-height of the outer disc in pc}  & ${\mu}_0$(2)\footnote{Face-on surface 
brightness of inner disc in mag arcsec$^{-2}$} & $R_D$(2)\footnote{Radial scale-length of inner disc in kpc} & $h_z$(2)\footnote{Vertical scale-height of the inner disc in pc} \\
\hline

IC5249       & 21.70 & 5.24  & 724 & 20.53 & 1.23 & 253 \\
UGC7321      & 21.73 & 2.39  & 436 & 19.94 & 1   & 134   \\
IC2233       & 21.67 & 2.16 & 338 & 20.82 & 0.81 &81 \\ \\

\hline
\end{tabular}}
\hfill{}
\label{tb:tablename}
\end{minipage}
\end{center}
\end{table}

\subsection{HI Surface Density and Rotation Curves} Modelling the de-projected surface density profile and the rotation 
 curve for perfectly edge-on galaxies may be tricky as common techniques of converting kinematics data into a
 disc rotation curve (tilted ring modelling, for example) fails in case of edge-on galaxies due to projection effects (See O'Brien et al. 2010b for a review). Like stellar photometry, the HI images for our sample galaxies were available
 in the literature, and the surface densities and rotation curves modelled, as discussed below. Our sample galaxies have almost 
symmetric rotation curves. In terms of surface density profiles, they are mildly lopsided except for IC2233, which indicates lopsidedness.
The HI disc was assumed to be optically thin in calculating the surface density values, as is indicated by the mildness of self-absorption
near the midplane in case of UGC7321 (Uson \& Matthews 2003). \\

\noindent \emph{IC5249:} The HI rotation curve and HI surface density for IC5249 have been taken from O'Brien, Freeman \& van der Kruit (2010c). 
The rotation curve could be obtained till $\sim$ 3.5 $R_D$(1), and the asymptotic rotational speed $v_{\rm{max}}$ 
is $\sim$ 112 kms$^{-1}$. The peak HI surface density ${\Sigma}_{HI,\rm{peak}}$ $\sim$ 8
M$_{\odot}$pc$^{-2}$ is the highest among our sample galaxies. It is a gas-rich galaxy as indicated by a high value of $M_{HI}/L_B$ $\sim$ 1.4 
where $M_{HI}$ is the total $HI$ mass and $L_B$ the total luminosity in the B-band. \\

\noindent \emph{UGC7321:} \noindent The HI rotation curve and HI surface density for UGC7321 has been taken from Uson \& Matthews (2003). 
UGC7321 has an asymptotic rotational velocity $v_{\rm{max}}$ $\sim$ 110 kms$^{-1}$, and the rotation curve could be obtained
 till 5 $R_D$(1). The peak HI surface density ${\Sigma}_{HI}$ $\sim$ 3 M$_{\odot}$pc$^{-2}$. It is a gas-rich and dark matter-dominated galaxy as indicated by a high value of $M_{HI}/L_B$ $\sim$ 1, and a 
large value of the dynamical mass-to-light ratio $M_{\rm{dyn}}/L_B$ $\sim$ 12, where $M_{\rm{dyn}}$ is the dynamical mass of the galaxy as 
estimated from the rotation curve. \\  

\noindent \emph{IC2233:} The HI rotation curve and HI surface density for IC2233 have been taken from Matthews \& Uson (2008)
. IC2233 has a $v_{\rm{max}}$ $\sim$ 85 kms$^{-1}$, the smallest among our sample galaxies, the rotation curve being measured till
 4 $R_D$(1). Unlike the other galaxies in the sample, IC2233 has a central hole, and hence the HI surface density peaks in the middle of the 
disc at ${\Sigma}_{HI}$  $\sim$ 3 M$_{\odot}$pc$^{-2}$. It is not so gas-rich with an $M_{HI}/L_B$ $\sim$ 0.62, but seem to be 
dark matter dominated ($M_{dyn}/M_{HI}$ $\sim$ 12).\\

\noindent {\bf Error-Bars on the Rotation Curves:} The original error-bars originally quoted on the above rotation curves were just fitting 
uncertainties 
and were therefore 
unreasonably small. We have attempted to use realistic error-bars on the rotational velocities by adding in quadrature the systemic 
errors to the fitting uncertainties, the systemic error being one-fourth of the velocity difference between the approaching and receding 
sides of the galaxy (See, for example, Uson \& Matthews 2003, de Blok et al. 2008). \\

\noindent The HI properties of our sample galaxies are presented
in Table 3.

\begin{table*}
\begin{center}
\begin{minipage}{150mm}
{\small
\hfill{}
\caption{HI 21cm properties}
\centering
\begin{tabular}{l|c|c|c|c|c}
\hline
Galaxy & ${\Sigma}_{HI,\rm{peak}}$\footnote{Peak HI surface density} & $M_{HI}$/$M_{L_B}$\footnote{Ratio of the total HI mass to blue luminosity} & $M_{\rm{dyn}}$/$M_{L_B}$\footnote{Ratio of the total dynamical mass to blue luminisoty} & $M_{\rm{dyn}}$/$M_{HI}$\footnote{Ratio of the total 
dynamical mass to total HI mass} & $V_{\rm{max}}$  \\
       & (M$_{\odot}$pc$^{-2}$)  &                    &                    &                    & (kms$^{-1}$)                \\
\hline

IC5249\footnote{Values quoted from O'Brien et al. 2010d}          & 8  & 1.4& -& -& 112\\
UGC7321\footnote{Values quoted from Uson \& Matthews 2003}        &  3 &  1&12& 31 & 110\\
IC2233\footnote{Values quoted from Matthews \& Uson 2008}         &  3 & 0.62& - & 14 & 85\\

\hline
\end{tabular}}
\hfill{}
\label{tb:tablename}
\end{minipage}
\end{center}
\end{table*}

\section{MASS MODELS}

\subsection{Stellar Component}

\noindent As discussed in \S 3.1, two-dimensional structural surface brightness decompositions at Spitzer 3.6 $\mu$m band have indicated
that the stars in our sample galaxies are distributed in two exponential discs with finite thickness, namely the \emph{inner} and the 
\emph{outer disc}. We use the GIPSY (Groningen Image Processing System) task ROTMOD to obtain the rotation curve 
due to the gravitational potential of each of these disc components. For each disc component, we give the face-on central surface brightness
(${\mu}_0$(1) or ${\mu}_0$(2) in L$_{\odot}$ pc$^{-2}$), radial scale-length ($R_D$(1) or $R_D$(2) in kpc) and vertical scale -length ($h_z$(1) or $h_z$(2) in kpc) as input parameters to ROTMOD, choosing the disc to be anti-truncated (\S 3.1). ROTMOD calculates the rotation
 curve of such a disc following the analytical formulations given, for example, in Casertano (1983). The net rotation curve due to the 
stellar disc is then  obtained by adding in quadrature the rotational velocities due to the \emph{inner} and the
 \emph{outer discs} respectively: $$V_{\rm{stars}}^2 = {V_{\rm{inner disc}}}^2 + {V_{\rm{outer disc}}}^2 \eqno(1)$$

\subsection{Neutral Gas}

\noindent The rotation curve due to the gas component $V_{\rm{gas}}(R)$ is obtained by assuming gas to be present in thin, concentric 
 rings. The HI surface density values were scaled by a factor of 1.4 to take into account the contribution 
from helium and metals. We have not taken into account the presence of molecular gas $H_2$ in our mass models. However, it may not be a 
dynamically significant component in superthin galaxies. Matthews et al. (2005) reported that the $H_2$ to $H_I$ mass ratio in late-type LSB 
spirals 
$\sim$ 10$^{-3}$. Therefore the contribution of $H_2$ can be ignored for the construction of mass models of LSB galaxies like the superthins,
 without incurring significant errors. \\

\noindent As with the case of the stellar component, we use the GIPSY task ROTMOD to model the rotation
 curve $V_{\rm{gas}}$ due to the gravitational potential of the neutral gas component. \\ 

\subsection{Dark Matter Halo}
\noindent We use the pseudo-Isothermal (PIS) and Navarro-Frank-White (NFW) density profiles to model the dark matter haloes
 of the superthin galaxies.

\subsubsection{Pseudo-Isothermal (PIS) Halo}

\noindent The density profile for a PIS dark matter halo is given by 
$${\rho}_{PIS} (R)= \frac{{\rho}_{0}} {1+(\frac{R}{R_{c}})^2} \eqno(2)$$
Here ${\rho_{0}}$ is the central core density and $R_{c}$ is the core radius of the halo. The rotational velocity at any radius R due to a 
PIS dark matter halo is given by
$$V_{DM}(R)= V_{\infty} \sqrt {1-(\frac{R_{c}}{R}) {\rm{arctan}}(\frac{R}{R_{c}})} \eqno(3)$$
where the asymptotic rotational velocity of the halo $V_{\infty}$ is
$$V_{\infty}=\sqrt {4 \pi G \rho_{0} R_{c}^2} \eqno(4)$$
(Binney \& Tremaine 1987) \\

\subsubsection{Navarro-Frenk-White (NFW) Dark Matter Halo}

\noindent The density profile for an NFW dark matter halo is given by
$$\frac{{\rho}_{NFW}}{{\rho}_{crit}}=\frac{{\delta}_{c}}{{\frac{R}{R_{s}}}{(1+{\frac{R}{R_{s}}})}^2} \eqno(5)$$
where $R_{s}$ is characteristic radius of the halo and ${\rho}_{crit}$ is the critical density of the universe and
$$ {\delta}_{c}=\frac{200}{3} \frac {c^3} {{\rm{ln}}(1+c)-\frac{c}{1+c}}  \eqno(6)$$
$c$ being the concentration parameter given by $$c=\frac{R_{200}}{R_{s}} \eqno(7)$$ and $R_{200}$ the radius where the average density of 
 the NFW halo is 200${{\rho}_{\rm{crit}}}$ and is roughly equal to the virial radius $R_{vir}$. The rotation curve for an NFW dark matter halo is 
given by $$ V_{DM}(R)=V_{200} \sqrt{\frac{1}{x} \frac{ {\rm{ln}}(1 + cx) - \frac{cx}{1 + cx} } { {\rm{ln}}(1 + c) - \frac{c}{1+c} }} \eqno(8)$$
$$x=\frac{R}{{R}_{200}} \eqno(9)$$ The velocity at $R_{200}$ is $$V_{200}= 0.73 R_{200} \eqno(10)$$  where $R_{200}$ is in kpc and $V_{200}$ 
is in kms$^{-1}$ (Navarro et al. 1996). \\ \\

\noindent Finally, the net rotation velocity $V_{R}$ at a radius $R$ is obtained by adding in quadrature the rotational velocity due to the 
gravitational potential 
due to each of the components: 
$$ {V(R)}^2 = {\Upsilon}* {V_{\rm{stars}}(R)}^2 + {V_{\rm{gas}}(R)}^2 + {V_{\rm{DM}}(R)}^2 \eqno(11)$$
where ${\Upsilon}*$ is the stellar mass-to-light ratio which is discussed in the next subsection.

\subsection{Stellar mass-to-light ratio ${\Upsilon}*$ } 

\noindent {\bf Constant ${\Upsilon}*$:} In order to convert the observed stellar surface brightness profiles at a given 
wavelength to their surface density profiles,
 we require their stellar mass-to-light ratio at that wavelength. Using stellar population synthesis models, Bell \& de Jong (2001)
obtained empirical relationships between optical colours of galactic discs and stellar mass-to-light ratios at various wavelengths.
However, such an empirical relationship for the Spitzer 3.6 $\mu$m band was lacking in the above study. Oh et al. (2008) derived
 the following empirical relationship between stellar mass-to-light ratios in the Spitzer 3.6 $\mu$m band (${{\Upsilon}*}^{3.6}$) and
in the K-band (${{\Upsilon}*}^{K}$) by constructing stellar
population  synthesis  models, with  various  sets  of  metallicity
and star-formation histories:
$$ {{\Upsilon}*}^{3.6} = B^{3.6} + {{\Upsilon}*}^{K} \times A^{3.6} \eqno(12)$$
where $A^{3.6}$ = -0.05 and  $B^{3.6}$ = 0.92. 
\noindent The empirical relation between optical colours and ${{\Upsilon}*}^{K}$ is given by
$$ {\rm{log}}_{10}({{\Upsilon}*}^{K}) = b_K \times \rm{Optical \space Colour} + a_K \eqno(13)$$
$a_K$ = -0.67 and $b_K$ = 0.42 for B - R colours with Scaled Salpeter IMF (Bell \& de Jong 2001).  
We assume ${\Upsilon}*$ to remain constant with radius in our mass models as colour gradients in LSB galaxies have been found to be small 
(de Blok et al. 2001). \\

\noindent The optical colours and mass-to-light ratios in the K and the Spitzer 3.6 $\mu$m band of our sample galaxies as used in the mass
models are summarised in Table 4. \\

\begin{table}
\begin{center}
\begin{minipage}{150mm}
{\small
\hfill{}
\caption{Optical colours and mass-to-light ratios}
\centering
\begin{tabular}{l|c|c|c}
\hline
Galaxy & B-R &
${{\Upsilon}*}^{K}$\footnote{Mass-to-light ratio in K band estimated using\\
 Bell \& Jong (2001)} &
${{\Upsilon}*}^{3.6 \mu m}$\footnote{Mass-to-light ratio at the Spitzer 3.6$\mu$ band estimated \\
using Bell \& Jong (2001) \& Oh et al. (2008)}  \\
\hline

IC5249       & 0.70\footnote{Quoted from Byun (1998)} & 0.34 & 0.31  \\
UGC7321      & 0.97\footnote{Quoted from Uson \& Matthews (2003)} & 0.45 & 0.42 \\
IC2233       & 0.67\footnote{Quoted from Matthews \& Uson (2008)} & 0.31 & 0.31 \\

\hline
\end{tabular}}
\hfill{}
\label{tb:tablename}
\end{minipage}
\end{center}
\end{table}

\noindent {\bf Free ${\Upsilon}*$:} Here, we do not fix the stellar mass-to-light ratio ${\Upsilon}*$ at any value determined by the population
 synthesis models, and allow it to vary along with the dark matter halo parameters (${\rho}_0$, $R_c$). \\

\noindent {\bf Maximum Disc:} The Maximum Disc model imposes a lower bound on the dark matter density distribution in the galaxy.
  In this model, the stellar disc is taken to dominate the underlying gravitational potential. This is realised by fixing  
 the mass-to-light ratio (${\Upsilon}*$) at an appropriate value such that a major contribution ($\sim$ 75 $\%$) to the peak of the rotation 
curve (R $\sim$ 2.2 $R_D$ or, 2.2 $R_D$(1) in our study as the \emph{outer disc} is the main disc component) is attributed to the 
gravitational potential of the stellar disc alone (Sackett 1997). 
The free parameters in this case ${\rho}_0$, $R_c$.\\

\noindent {\bf Minimum Disc:} Similarly, the Minimum Disc model sets an upper bound on the dark matter density. 
In this case, the disc contribution to the underlying gravitational potential is taken to be zero, and the observed rotation curve is
attributed entirely to the potential of the dark matter halo. The free parameters in this case are again ${\rho}_0$, $R_c$. \\

\subsection{Modified Newtonian Dynamics (MOND)}
The rotation curves of low surface brightness galaxies provide a unique data set with which to test
alternative theories of gravitation like the Modified Newtonian Dynamics (MOND). \\

\noindent The net rotational velocity V(R) under MOND is given by ${V(R)}^2 = $
$$ \frac{1}{2} ({\Upsilon}* {V_{\rm{stars}}(R)}^2 + {V_{\rm{gas}}(R)}^2)   
(1 + \sqrt {1 + {(\frac {2RA}{ {V_{\rm{stars}}(R)}^2 + {V_{\rm{gas}}(R)}^2 }^2})}  \eqno(14)$$
(Milgrom 1983). Here the two free parameters are ${\Upsilon}*$ and A, the mass-to-light ratio and the acceleration per unit length (in units
 of kms$^{-1}$kpc$^{-1}$) respectively. \\

\noindent Finally, we use these stellar rotation curve ($V_{\rm{stars}}$) and gas rotation 
curves ($V_{\rm{gas}}$) as derived using the GIPSY task ROTMOD and the observed rotation curve task as inputs to the task ROTMAS, an 
interactive routine to fit different components (stars, gas and dark matter/MOND) for different mass models to the observed rotation curve.  

\section{Results \& Discussion} 

\subsection{PIS and NFW DM Haloes} We present here mass models for each of our sample
galaxies using PIS and NFW DM density profiles. We compare our best-fitting model rotation curves with the observed
one for different cases as discussed in \S 4.3: [Panel 1] the
Constant ${\Upsilon}*$ case, [Panel 2] the Free ${\Upsilon}*$ case, [Panel 3] the Maximum Disc, and [Panel 4] the Minimum Disc case
respectively. In each plot, we also indicate the rotation curve decompositions i.e., contribution of the stars, gas and dark matter halo
to the model rotation curve. However, we do not present the cases where the best-fitting mass models are not physically meaningful.\\

\subsubsection{IC5249} We present model rotation curves for IC5249 with a PIS and NFW DM Halo  
in Figures 2 and 3 respectively. All models with a PIS DM halo except for the Maximum Disc case, 
give good fits to the observed rotation curve except 
for slight mismatches at the outermost radii; an underestimation of the observed values at $R$ $\sim$ 12.5 kpc, and an 
over-estimation at $R$ $\sim$ 17 kpc. With respect to the Constant ${\Upsilon}*$ and Minimum Disc cases, IC5249 has the smallest central 
core density ${\rho}_0$ and the largest core radius $R_c$ among our sample galaxies, 
and the most \emph{compact} dark matter halo after UGC7321.
With the Free ${\Upsilon}*$ case, the best-fitting values of all the parameters including 
the ${\Upsilon}*$ value, 
closely match those of the Constant ${\Upsilon}*$ case within error-bars, confirming further the viability of the 
Constant ${\Upsilon}*$ case with the observed rotation curve.
The poor fit with the Maximum Disc model means that the disc dynamics is not strongly regulated by the gravitational potential of the 
stellar disc, which, indirectly implies dark matter dominance at all radii. This is further confirmed by the rotation curve decompositions 
which clearly show the dominance of 
the rotational velocity due to the gravitational potential of the dark matter halo at all radii. \\

\noindent Similarly, an NFW halo with the Constant ${\Upsilon}*$ and the Minimum Disc cases, give good fits to the 
rotation curve of UGC7321, the quality of the fits being comparable with those of the PIS DM Halo models. 
As with the PIS DM Halo, the Maximum Disc case with an NFW Halo does not comply with the observed rotation curve, yielding a negative 
$c$ value and an unusually high $V_{200}$ for the best-fitting case.

\begin{figure*}
\begin{center}
\begin{tabular}{cccc}
\resizebox{40mm}{!}{\includegraphics{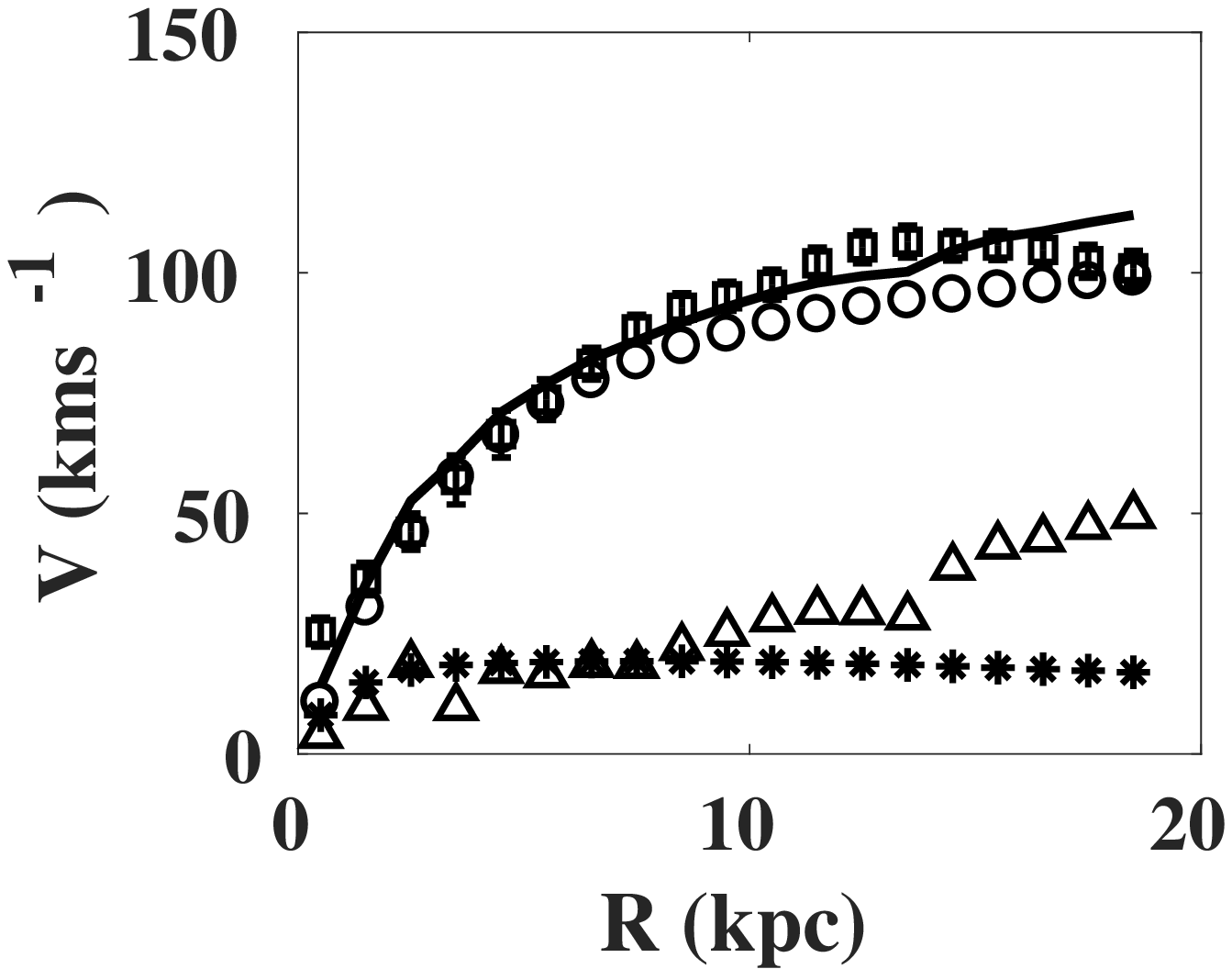}} &
\resizebox{40mm}{!}{\includegraphics{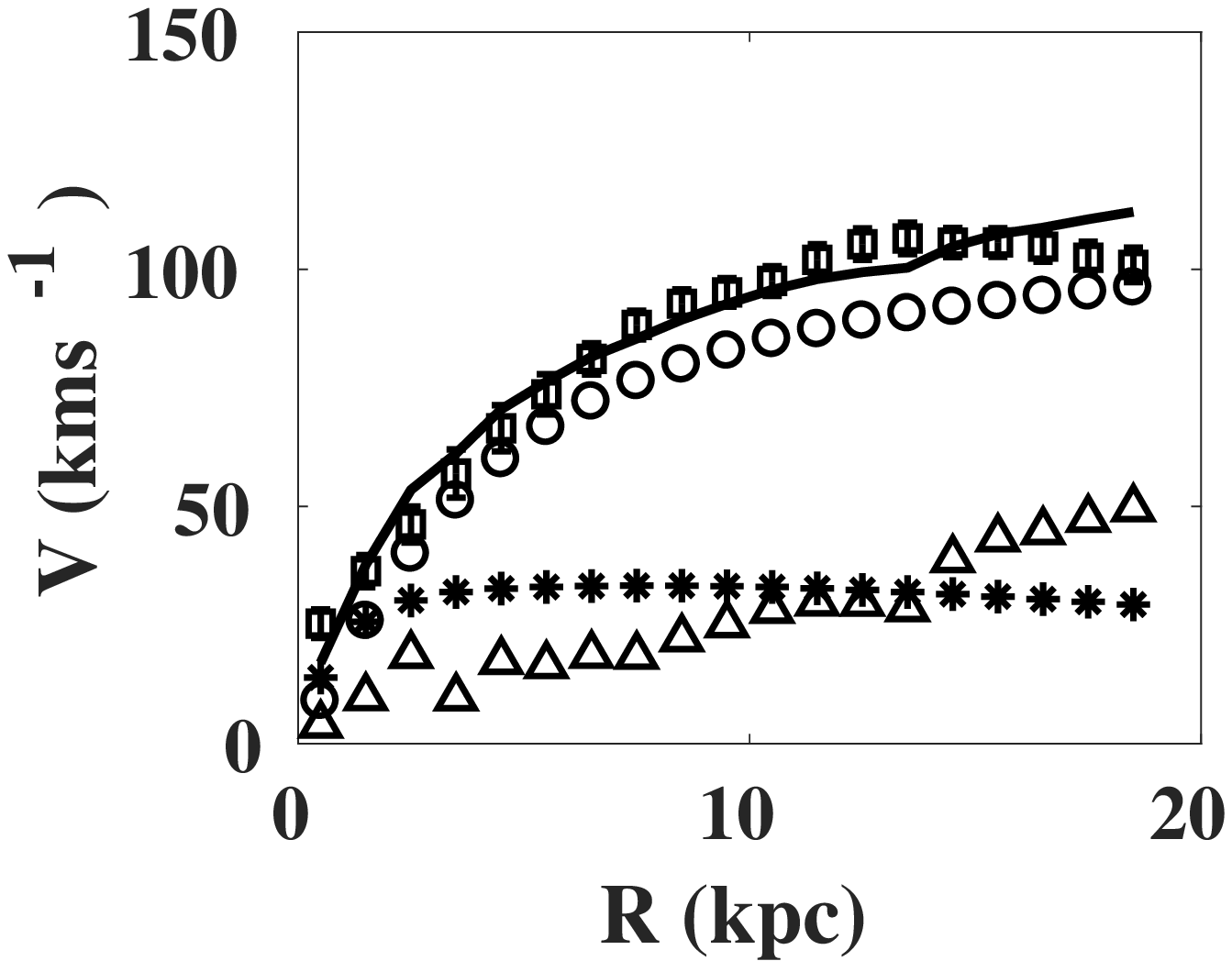}} &
\resizebox{40mm}{!}{\includegraphics{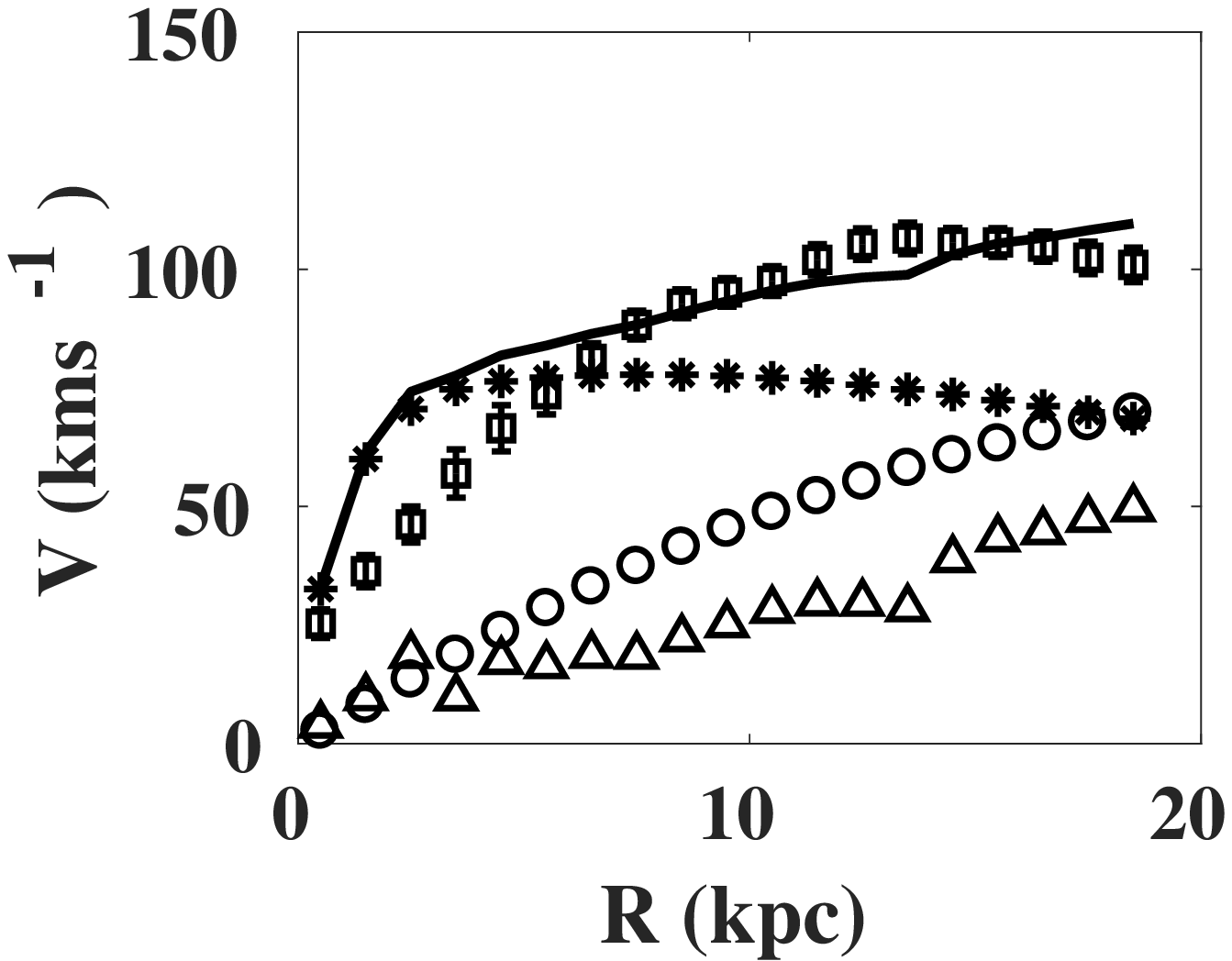}} &
\resizebox{40mm}{!}{\includegraphics{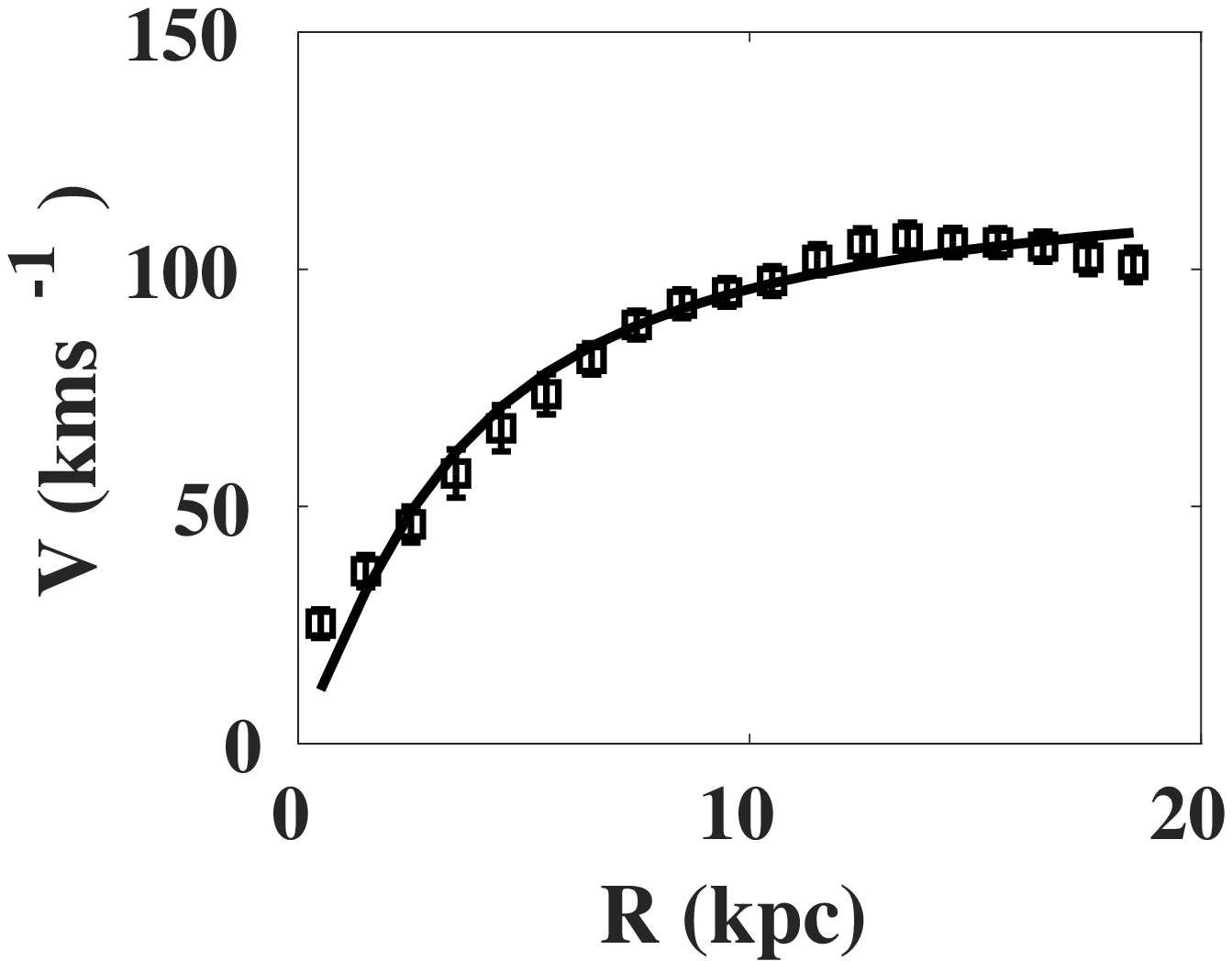}} \\
\end{tabular}
\end{center}
\caption{Modelling HI rotation curve of the superthin galaxy IC5249 with a PIS dark matter density
profile: Panel [1] A constant ${\Upsilon}*$
case as predicted by stellar population synthesis models, Panel [2] A free ${\Upsilon}*$ case, Panel [3] a Maximum disc case, and Panel [4]
a mimimum disc case. The \emph{stars} indicate the rotation curve due to the stellar disc alone, the \emph{triangles} the rotation curve due
to the gas disc, the \emph{open circles} that due to the dark matter halo, the \emph{solid line} the best-fitting model rotation curve and
the \emph{squares} the observed rotation curve with error-bars. }
\label{fig:pis_5249}
\end{figure*}

\begin{figure*}
\begin{center}
\begin{tabular}{cccc}
\resizebox{40mm}{!}{\includegraphics{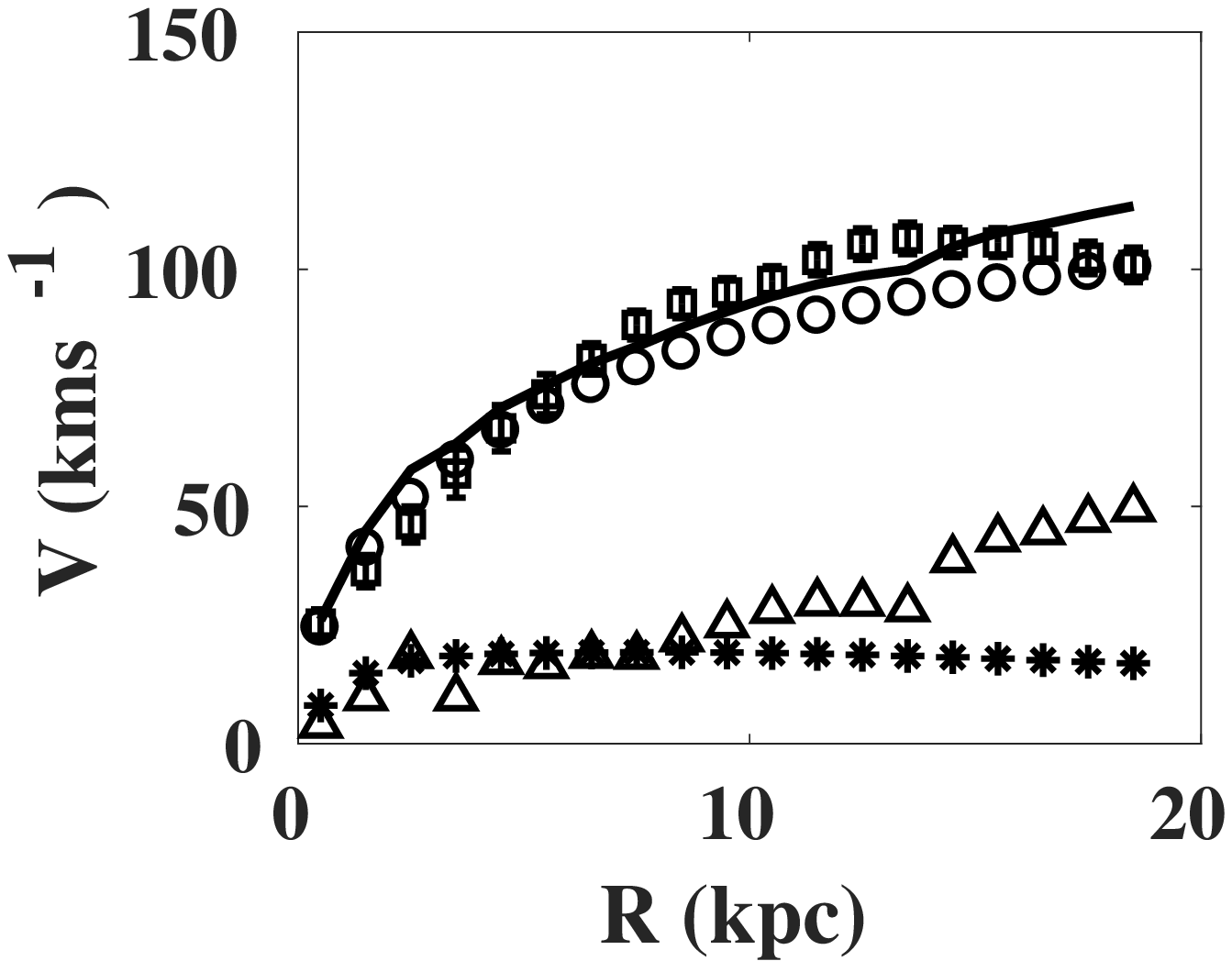}} &
\resizebox{40mm}{!}{\includegraphics{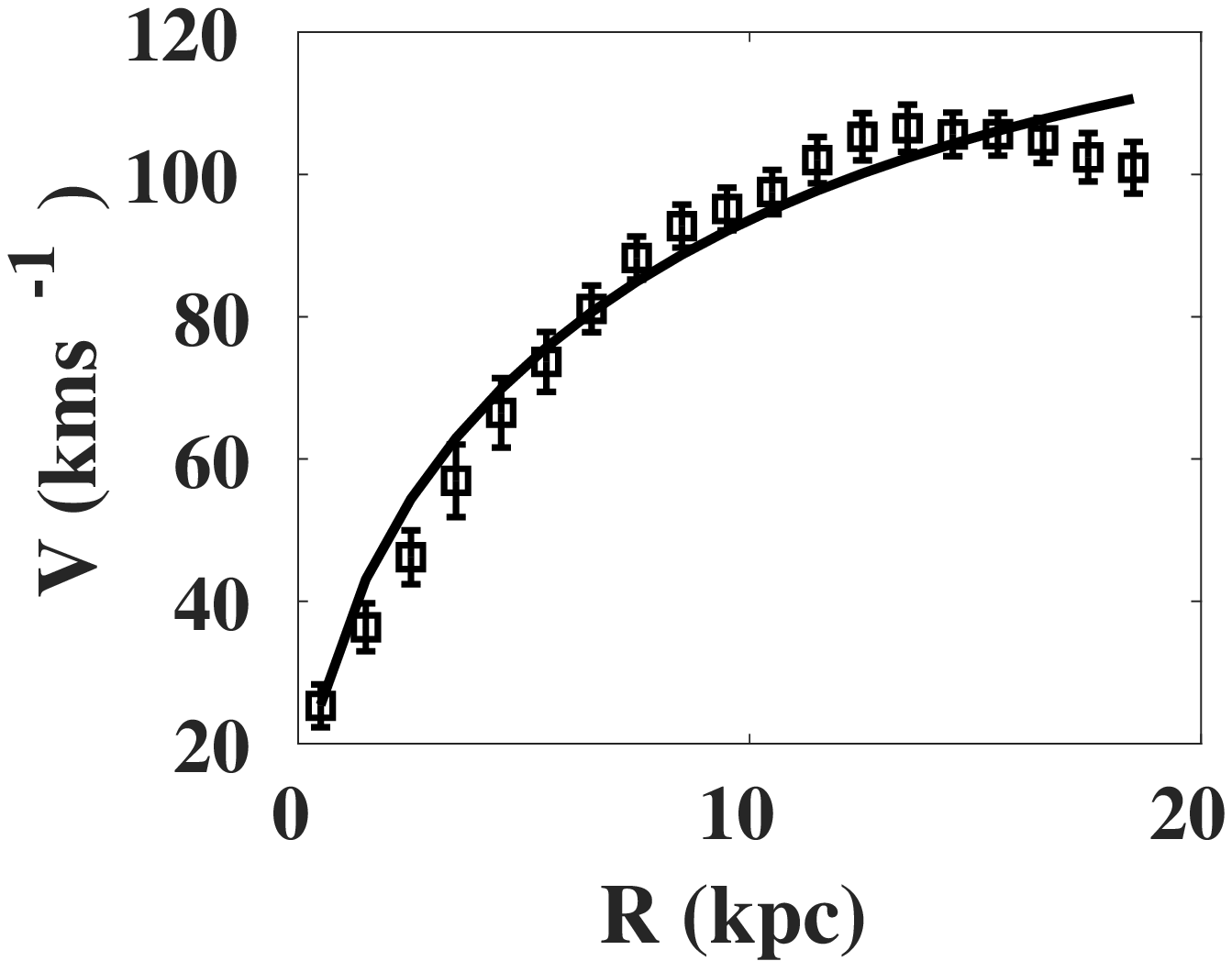}} \\
\end{tabular}
\end{center}
\caption{Modelling HI rotation curve of the superthin galaxy IC5249 with an NFW dark matter density profile: Panel [1] A constant ${\Upsilon}*$
case as predicted by stellar population synthesis models and Panel [2]
a mimimum disc case. The \emph{stars} indicate the rotation curve due to the stellar disc alone, the \emph{triangles} the rotation curve due
to the gas disc, the \emph{open circles} that due to the dark matter halo, the \emph{solid line} the best-fitting model rotation curve and
the \emph{squares} the observed rotation curve with error-bars. }
\label{fig:nfw_5249}
\end{figure*}

\subsubsection{UGC7321}  We present model rotation curves for UGC7321 with a PIS and NFW DM Halo in Figures 4 and 5 respectively. 
As with IC5249, all models with a PIS DM halo except for the Maximum Disc case,
give good fits to the observed rotation curve, possibly the best among all our sample galaxies. 
With respect to both the Constant ${\Upsilon}*$ and the Minimum Disc cases, 
UGC7321 has the highest value of ${\rho}_0$, the
smallest $R_c$ and the most compact dark matter halo i.e., the smallest $R_c/R_D$ value among our sample galaxies. 
In other words, UGC7321 has the densest and the most compact dark matter halo.
Our best-fitting case for a Constant ${\Upsilon}*$
 model predicts ${\rho}_0$ $\sim$ 0.14 M$_{\odot}$pc$^{-3}$ and $R_c$ $\sim$ 1.27 kpc, which do not match with the results
obtained in earlier studies.
Using the joint constraints of the HI rotation curve and 
the HI scale-height data, Banerjee et al. (2010) found that the
UGC 7321 has ${\rho}_0$ $\sim$ 0.039 - 0.057 M$_{\odot}$ pc$^{-2}$ and $R_c$ $\sim$ 2.5 - 2.9 kpc.  
In a similar study, O'Brien et al. (2010d) found a PIS DM halo with a ${\rho}_0$ $\sim$ 0.73 
 M$_{\odot}$pc$^{-3}$ and $R_c$ $\sim$ 0.52 kpc for UGC7321.
This mismatch may be possibly explained by the fact that different stellar surface brightness profiles used in these
 studies resulted in different dynamical contribution of the stellar disc in the mass models.
While the optical photometry used by Banerjee et al. (2010) and O'Brien et al. (2010d) indicated a low surface brightness single 
exponential stellar disc, Spitzer 3.6 $\mu$m images used in the current work gives a high surface brightness stellar disc, which is 
double exponential in nature (\S 3.1). The presence of this additional central component in the Spitzer
images may be a possible reason behind obtaining a best-fitting mass model having a dark matter halo with a smaller central core density and a larger core radius in the current work, compared to those found in the earlier studies. Besides, Banerjee et al. (2010) 
neglected the radial term in the Poisson Equation (under the
assumption of a flat rotation curve) in their calculations, which is, however, a reasonable assumption in the outer galaxy only. \\

\noindent With the NFW halo model, the Constant ${\Upsilon}*$ and the Minimum Disc cases, give good fits to the observed rotation curve, 
 with UGC7321 having the most concentrated and the smallest NFW halo among our sample galaxies. The Free ${\Upsilon}*$ and the Minimum Disc
 cases, as in IC5249, result in fits which are not physically meaningful. 

\begin{figure*}
\begin{center}
\begin{tabular}{cccc}
\resizebox{40mm}{!}{\includegraphics{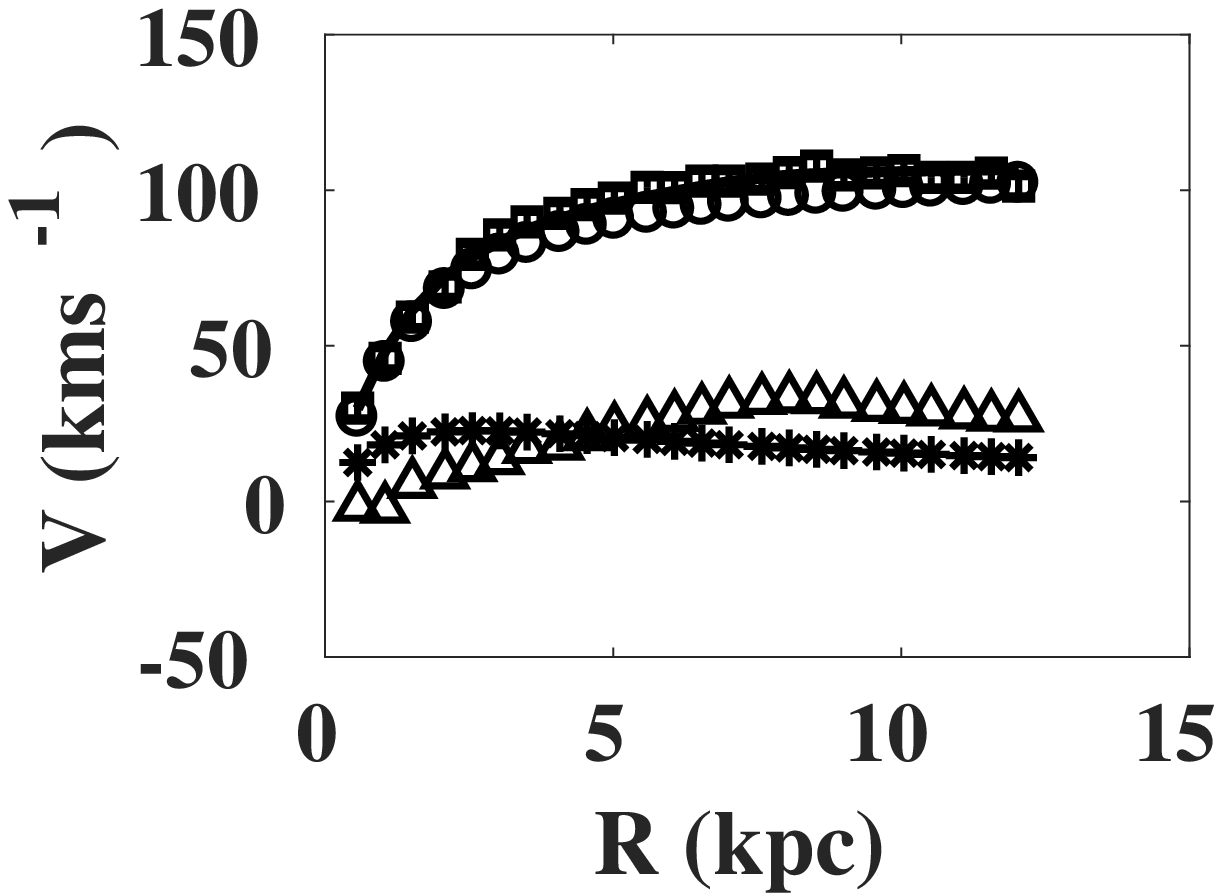}} &
\resizebox{40mm}{!}{\includegraphics{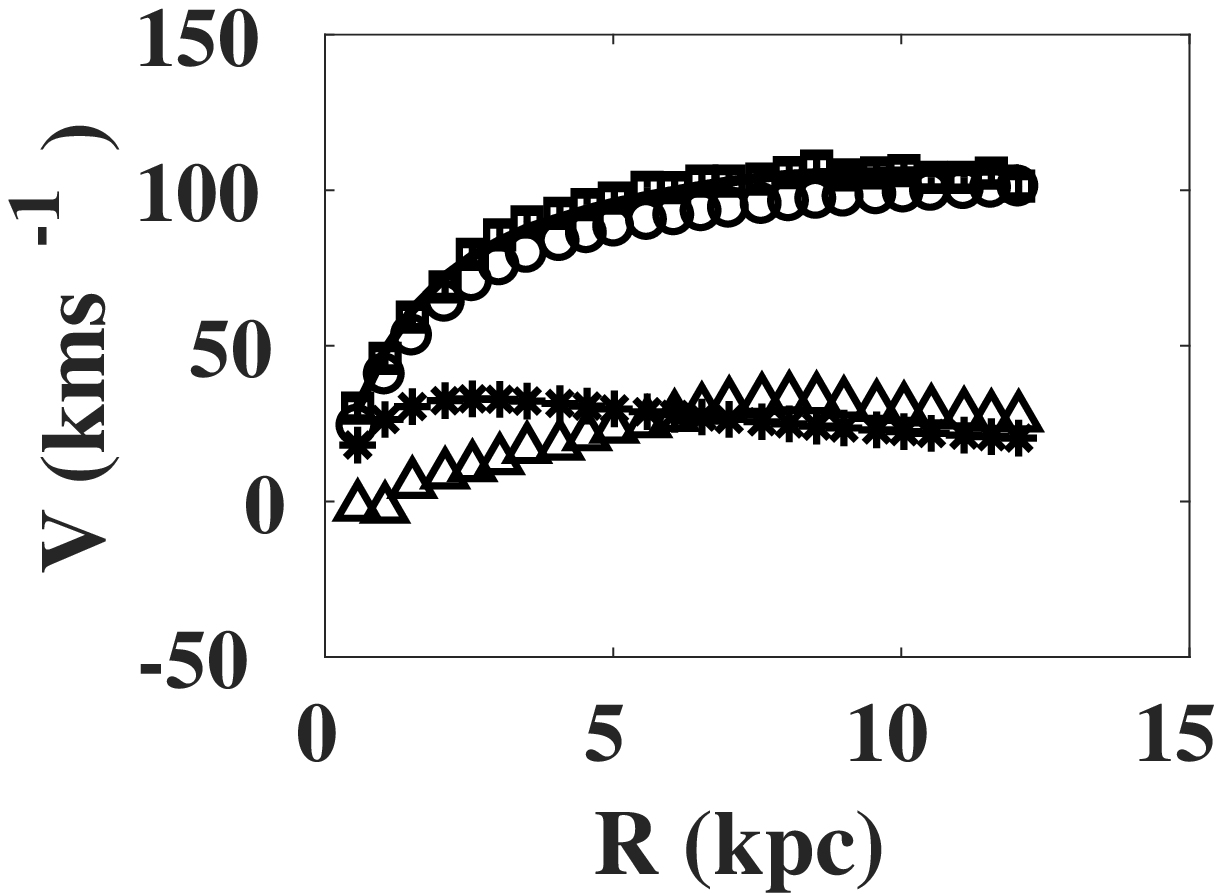}} &
\resizebox{40mm}{!}{\includegraphics{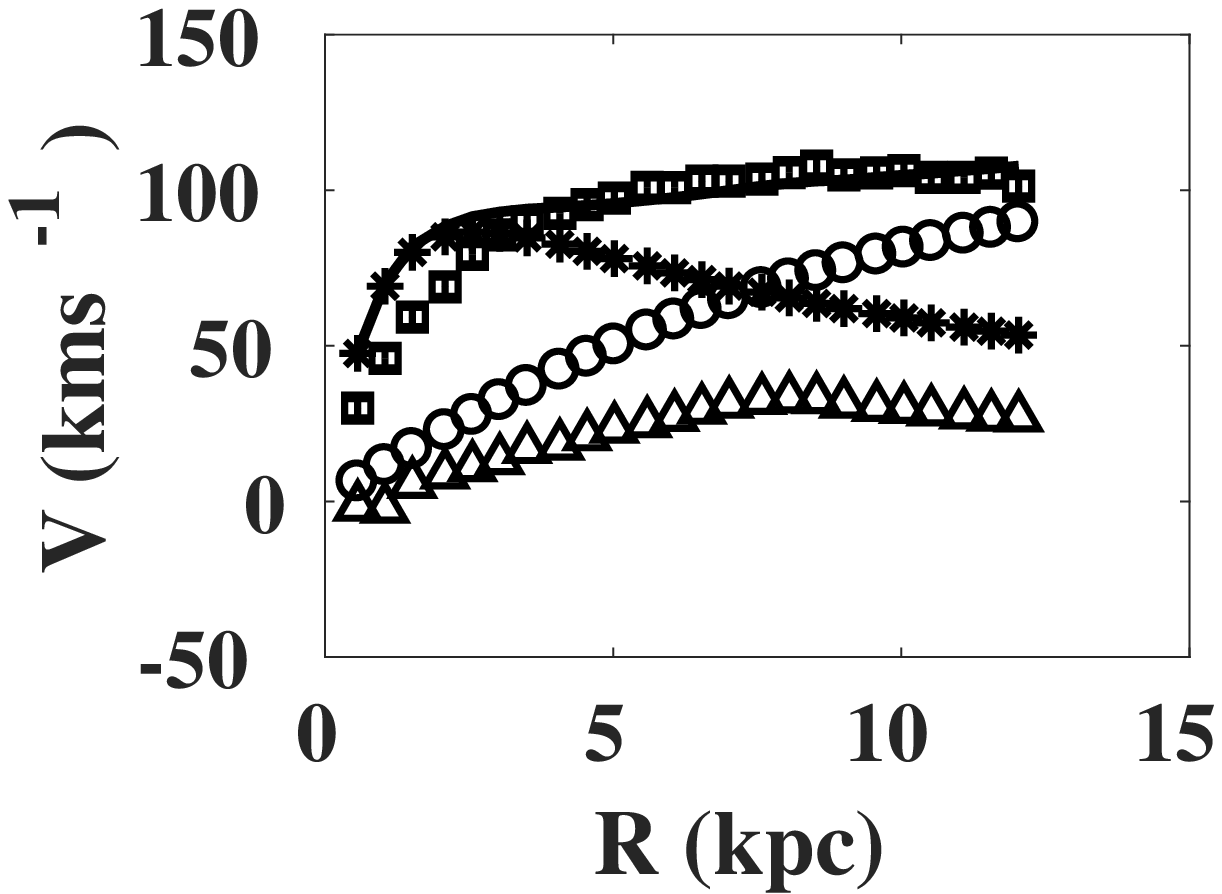}} &
\resizebox{40mm}{!}{\includegraphics{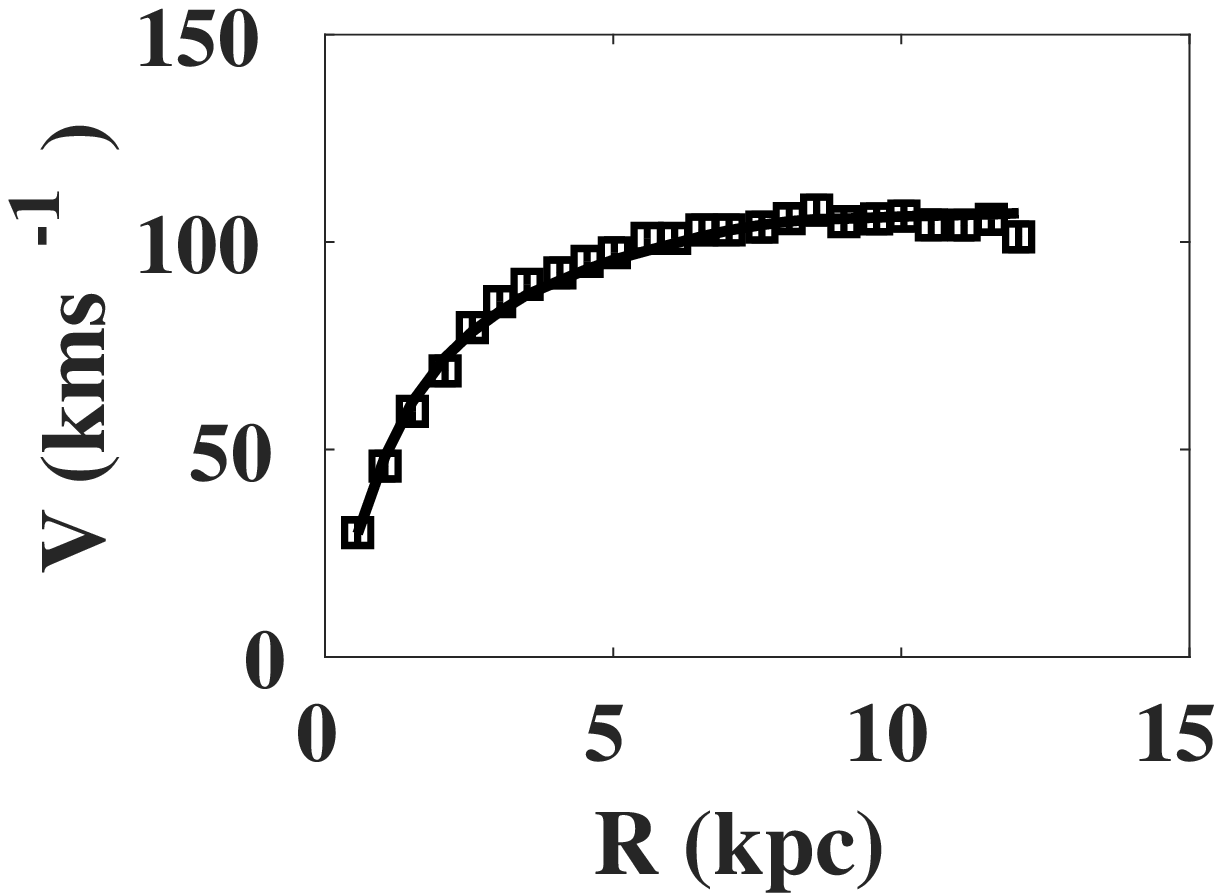}} \\
\end{tabular}
\end{center}
\caption{Modelling HI rotation curve of the superthin galaxy UGC7321 with a PIS dark matter density
profile: Panel [1] A constant ${\Upsilon}*$
case as predicted by stellar population synthesis models, Panel [2] A free ${\Upsilon}*$ case, Panel [3] a Maximum disc case, and Panel [4]
a mimimum disc case. The \emph{stars} indicate the rotation curve due to the stellar disc alone, the \emph{triangles} the rotation curve due
to the gas disc, the \emph{open circles} that due to the dark matter halo, the \emph{solid line} the best-fitting model rotation curve and
the \emph{squares} the observed rotation curve with error-bars. }
\label{fig:pis_7321}
\end{figure*}

\begin{figure*}
\begin{center}
\begin{tabular}{cccc}
\resizebox{40mm}{!}{\includegraphics{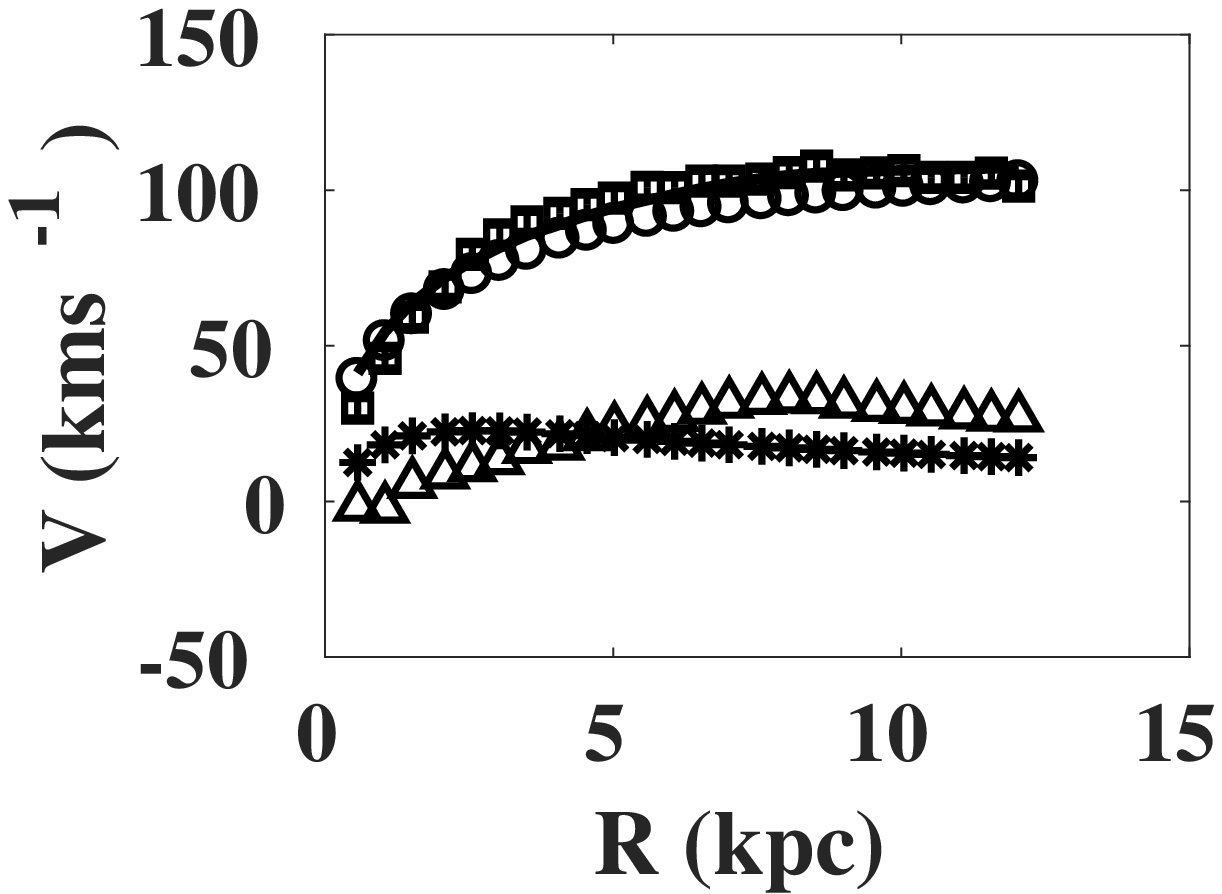}} &
\resizebox{40mm}{!}{\includegraphics{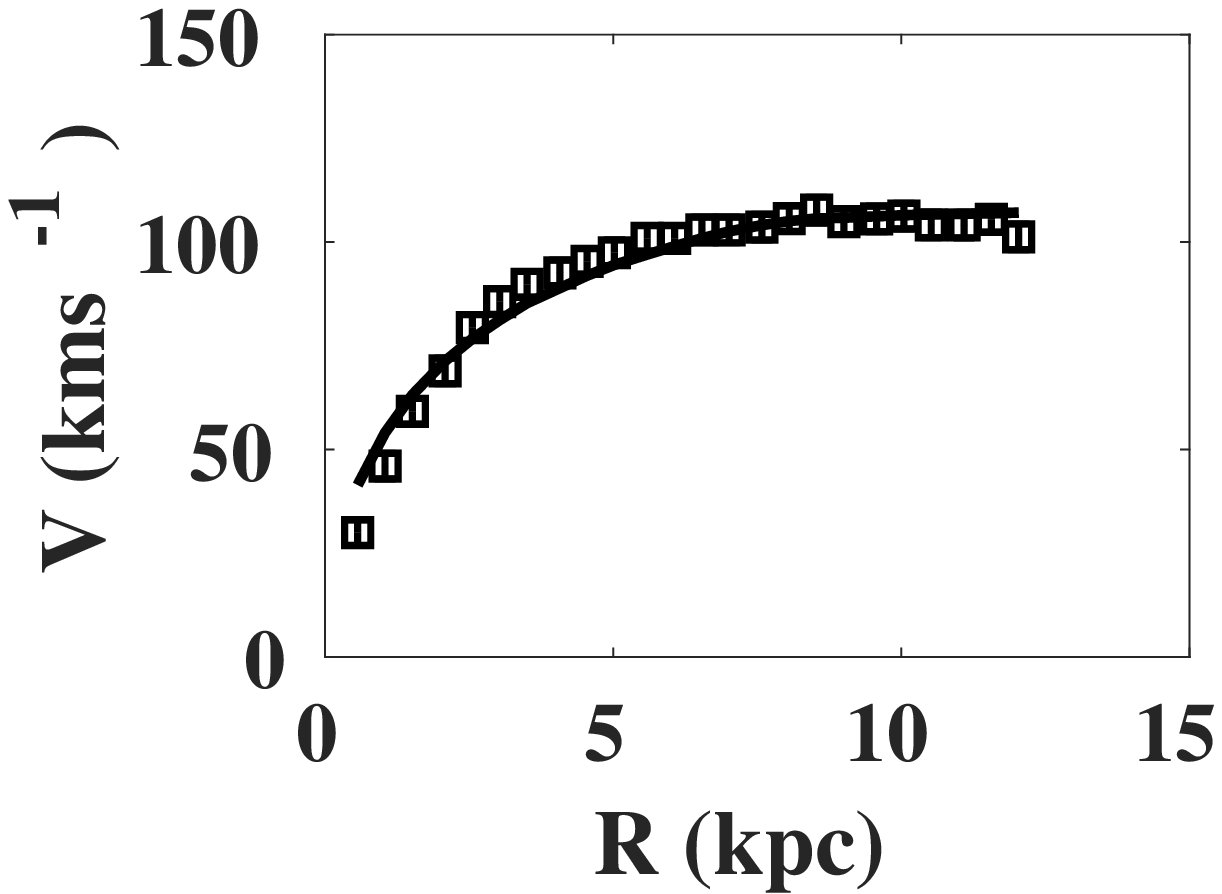}} \\
\end{tabular}
\end{center}
\caption{Modelling HI rotation curve of the superthin galaxy UGC7321 with an NFW dark matter density
profile: Panel [1] A constant ${\Upsilon}*$
case as predicted by stellar population synthesis models, and Panel [2]
a mimimum disc case. The \emph{stars} indicate the rotation curve due to the stellar disc alone, the \emph{triangles} the rotation curve due
to the gas disc, the \emph{open circles} that due to the dark matter halo, the \emph{solid line} the best-fitting model rotation curve and
the \emph{squares} the observed rotation curve with error-bars. }
\label{fig:nfw_7321}
\end{figure*}

\begin{figure*}
\begin{center}
\begin{tabular}{cccc}
\resizebox{40mm}{!}{\includegraphics{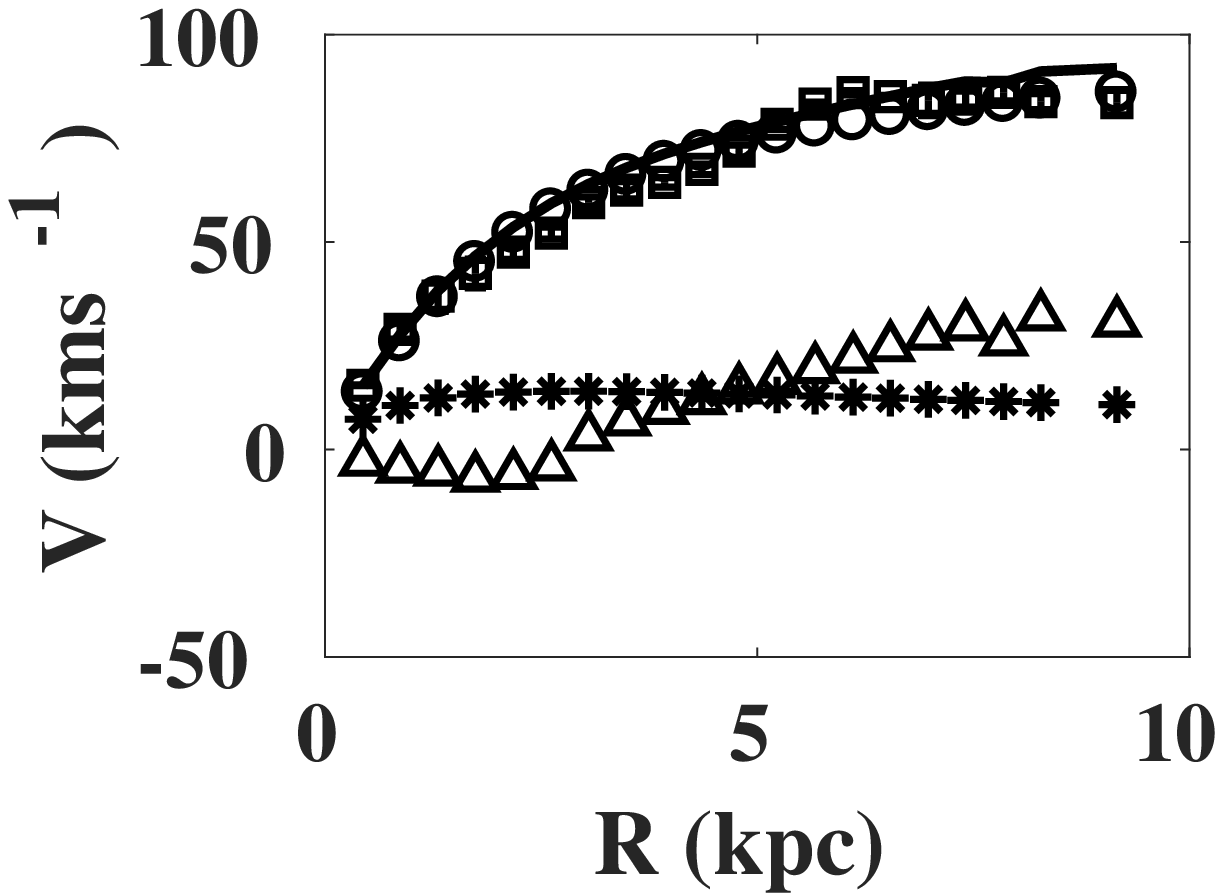}} &
\resizebox{40mm}{!}{\includegraphics{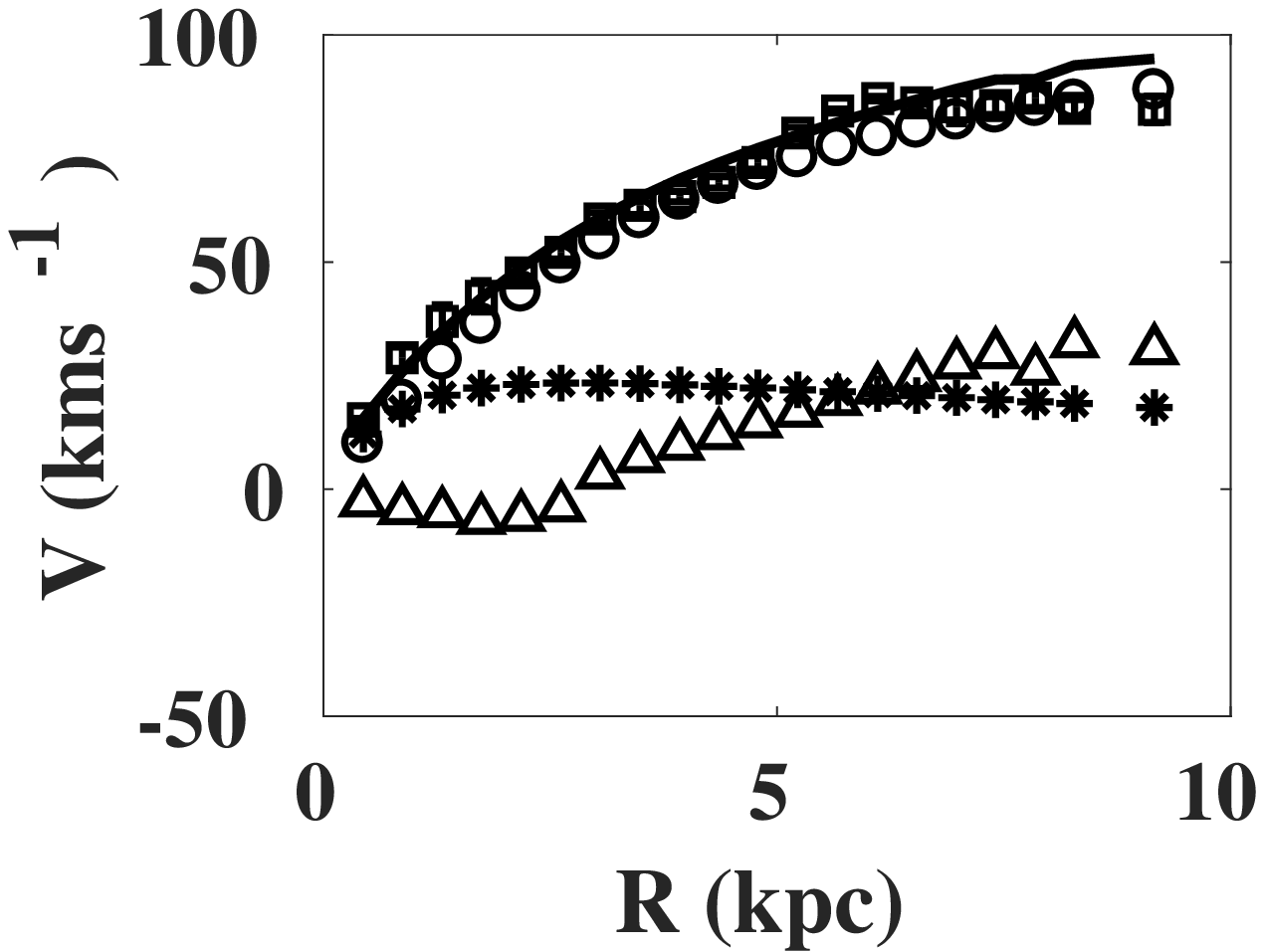}} &
\resizebox{40mm}{!}{\includegraphics{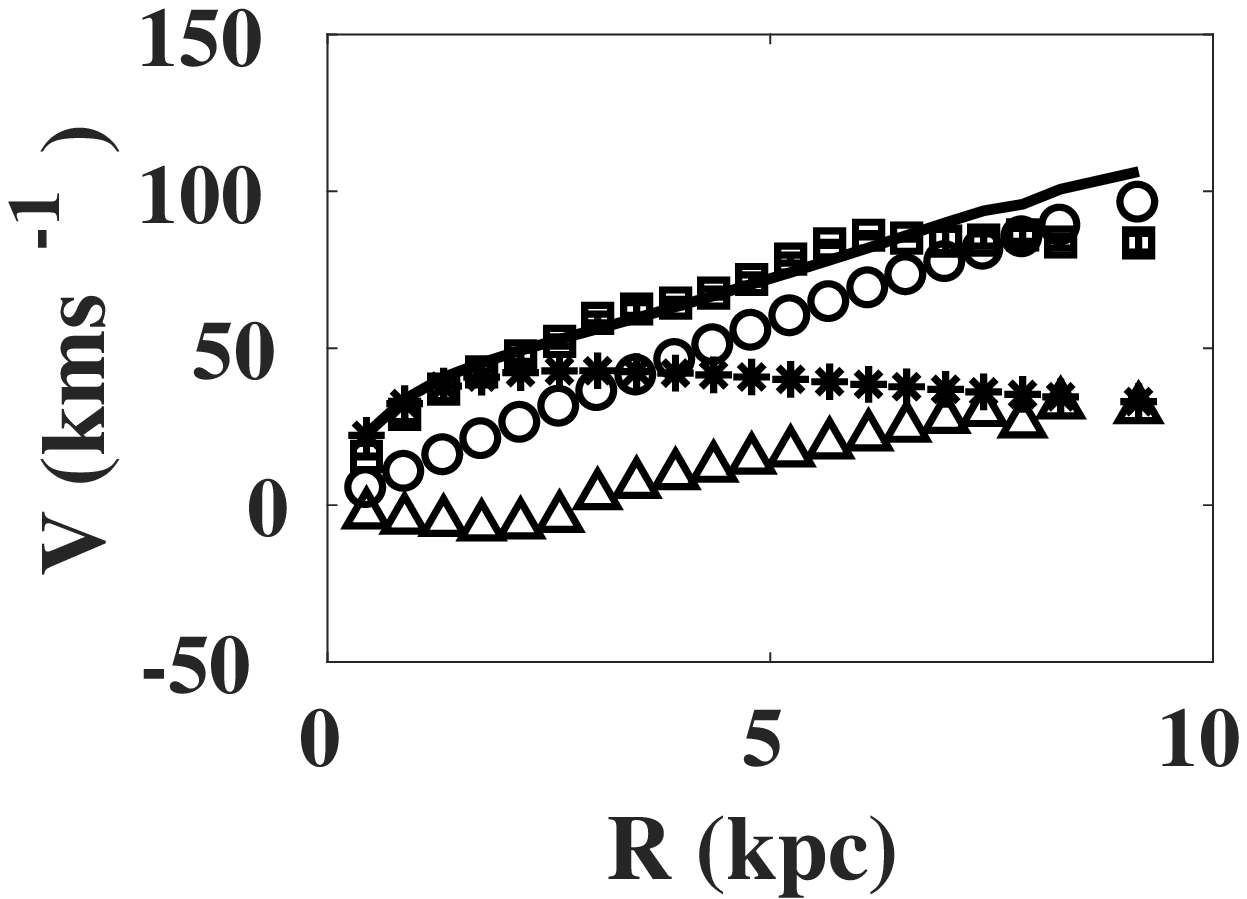}} &
\resizebox{40mm}{!}{\includegraphics{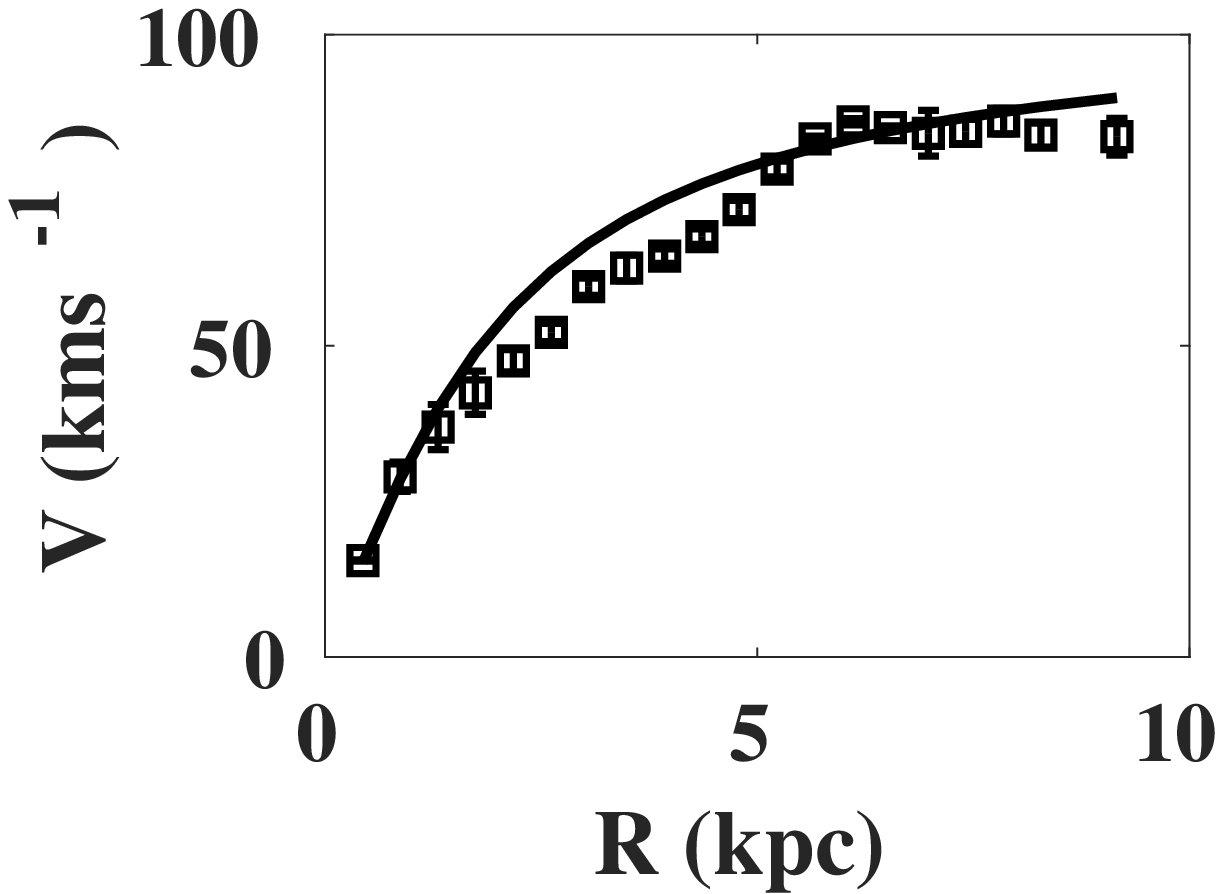}} \\
\end{tabular}
\end{center}
\caption{Modelling HI rotation curve of the superthin galaxy IC2233 with a PIS dark matter density
profile: Panel [1] A constant ${\Upsilon}*$
case as predicted by stellar population synthesis models, Panel [2] A free ${\Upsilon}*$ case, Panel [3] a Maximum disc case, and Panel [4]
a mimimum disc case. The \emph{stars} indicate the rotation curve due to the stellar disc alone, the \emph{triangles} the rotation curve due
to the gas disc, the \emph{open circles} that due to the dark matter halo, the \emph{solid line} the best-fitting model rotation curve and
the \emph{squares} the observed rotation curve with error-bars. }
\label{fig:pis_2233}
\end{figure*}

\subsubsection{IC2233}  We present mass models for IC2233 with a PIS DM halo in Figure 6. 
We note that the modelled gas rotation velocity within the inner disc indicate negative values at some radii.
We emphasise that the negative values of modelled gas rotation velocity within the inner disc simply indicate velocities in the 
outward/radial direction, and is characteristic of discs with central depressions as in IC2233 (See \S 5.1 of de Blok et al. 2008 for a 
discussion).  \\ 

\noindent As with the other galaxies in the sample, except for the Maximum disc case, the models with a PIS DM Halo give reasonable fits to the 
observed rotation curve, although the high values of the reduced 
chi-square indicate worse fits for the same. This may attributed to small, possibly under-estimated values of the error-bars obtained for the 
observed rotation curve (See \S 3.2), which happen to be an order of magnitude smaller than those of the other sample galaxies. \\ 

\begin{table*}
\begin{center}
\begin{minipage}{150mm}
{\small
\hfill{}
\caption{Results: Mass Models with PIS DM Halo}
\centering
\begin{tabular}{l|l|l|l|l|l|l}
                                                                                  
\hline
Galaxy & Model       &${\rho}_{0}$ \footnote{Central core density of the PIS DM Halo}                 & $Rc$ \footnote{Core radius of the PIS DM Halo}    & $Rc/R_D(1)$ \footnote{Ratio of the core radius to the exponential stellar disc scale-length}& ${\Upsilon}*$ \footnote{Stellar mass-to-light ratio}&  ${{{\chi}}_r}^2$  \footnote{Reduced chi-square corresponding to this best-fitting model}                          \\

       &       &($10^{-3} M_{\odot} \rm{pc}^{-3}$) & (kpc) &          &       &                                           \\

\hline

IC5249 & Constant ${\Upsilon}*$ &26.04$\pm$3.56 &2.99$\pm$0.27 & 0.57$\pm$0.05 &0.31 &2.21\\
       & Free ${\Upsilon}*$     &18.13$\pm$6.56  & 3.58$\pm$0.71& 0.68$\pm$0.14& 0.93$\pm$0.50 & 2.12\\
       & Maximum disc &1.65$\pm$0.91  & 12.78$\pm$9.18 &2.44$\pm$1.75  &5.06 &10.32 \\
       & Minimum disc & 28.89$\pm$2.75 & 3.11$\pm$0.19 & 0.59$\pm$0.04& 0.& 1.59\\ \\

UGC7321 & Constant ${\Upsilon}*$ & 140.23$\pm$9.37&1.27$\pm$0.05 &0.53$\pm$0.02 &0.42 &0.47 \\
       & Free   ${\Upsilon}*$     &  109.37$\pm$133.39  & 1.44$\pm$0.9 &0.60$\pm$0.37 & 0.88$\pm$2.05&0.47\\
       & Maximum disc &6.89$\pm$2.81 & 7.75$\pm$3.78& 3.24$\pm$1.58 &6.07 &10.41  \\
       & Minimum disc & 173.63$\pm$11.7 & 1.15$\pm$0.05 & 0.48$\pm$0.02& 0 & 0.51 \\ \\

IC2233 & Constant ${\Upsilon}*$ & 55.83$\pm$3.14 & 1.83$\pm$0.07 &0.85$\pm$0.03 & 0.31 & 11.8\\
       & Free   ${\Upsilon}*$     & 29.13$\pm$6.07&2.82$\pm$0.44 &1.31$\pm$0.20 & 0.86$\pm$0.14 &8.8 \\
       & Maximum disc &8.09$\pm$5.17  &12.12$\pm$28.57 &5.61$\pm$13.23 &2.88 &$\sim$310 \\
       & Minimum disc &  68.73$\pm$3.79& 1.71$\pm$0.06& 0.79$\pm$0.03& 0.& 16.8\\ \\

\hline
\end{tabular}}
\hfill{}

\label{tb:tablename}
\end{minipage}
\end{center}
\end{table*}

\begin{table*}
\begin{center}
\begin{minipage}{150mm}
{\small
\hfill{}
\caption{Results: Mass Models with NFW DM Halo}
\centering
\begin{tabular}{l|l|l|l|l|l|l|l}
                                                                                
\hline
 Galaxy      &   Model    &c \footnote{Concentration parameter of the NFW DM Halo}                 & $R_{200}$ \footnote{Radius where the average density of the NFW DM Halo is 200 ${\rho}_{crit}$}   &  $V_{200}$ \footnote{Rotational velocity at $R_{200}$}    & $\frac{V_{max}}{V_{200}}$ \footnote{Ratio of the maximum rotational velocity to the velocity at $R_{200}$}& ${\Upsilon}*$ \footnote{Stellar mass-to-light ratio}& ${{\chi}_r}^2$ \footnote{Reduced chi-square corresponding to this best-fitting model}                           \\

       &       &                   & (kpc)      &   (kms$^{-1}$) &   &        &                                                  \\ \\
\hline

IC5249 & Constant ${\Upsilon}*$ &3.48$\pm$0.48 & 78.65$\pm$5.28& 57.41 $\pm$ 3.85 &1.95 $\pm$ 0.13 &0.31&3.26 \\
       & Free  ${\Upsilon}*$ & 8.74$\pm$2.47 & 76.56$\pm$2.06 & 55.89$\pm$1.50 & 2.00$\pm$0.05 & -4.98$\pm$2.63&2.13 \\
       & Maximum disc & -0.43$\pm$ -  & 244.58$\pm$ & -  & - & 5.06 &12.56  \\
       & Minimum disc &  3.27$\pm$0.31 & 90.68$\pm$4.62 & 66.19$\pm$3.37 & 1.69$\pm$0.08 & 0 & 1.88 \\ \\

UGC7321 & Constant ${\Upsilon}*$ &7.80$\pm$0.60 &66.75$\pm$2.50& 48.73$\pm$1.83 &2.26$\pm$0.08 &0.42&1.87\\
       & Free  ${\Upsilon}*$     & 9.25$\pm$3.68& 64.56$\pm$5.42&47.13$\pm$3.96  &2.33$\pm$0.19 & -0.32$\pm$1.66& 1.88\\
       & Maximum disc & -0.43$\pm$ & 668$\pm$ -& - & -&6.07 &14.87 \\
       & Minimum disc &8.63$\pm$0.59 &65.38$\pm$2.14 &47.73$\pm$1.56 &2.30$\pm$0.07 &0.& 1.81\\ \\

IC2233  & Constant ${\Upsilon}*$ &0.15$\pm$7.27  &371.00$\pm$ -   & - &-  &0.31  & $\sim$ 213 \\
        & Free  ${\Upsilon}*$ & 5.43$\pm$4.45 & 81.33$\pm$30.42 & -& -&-3.96$\pm$3.52 & 20.49\\
        & Maximum disc& -0.43$\pm$ - & 614.04$\pm$ - & -& -& 2.88 & $\sim$ 877 \\
        & Minimum disc & -0.44$\pm$ - &1097.32$\pm$-  & -& -& 0. & $\sim$ 3756\\ \\
\hline
\end{tabular}}
\hfill{}
\label{tb:tablename}
\end{minipage}
\end{center}
\end{table*}

\begin{table*}
\begin{center}
\begin{minipage}{150mm}
{\small
\hfill{}
\caption{Results: Mass Models with MOND}
\centering
\begin{tabular}{l|l|l|l}
\hline
Galaxy       &A \footnote{Acceleration per unit length}                 & ${\Upsilon}*$ \footnote{Stellar mass-to-light ratio}   & ${{{\chi}}_r}^2$  \footnote{Reduced chi-square corresponding to this best-fitting model}                          \\

       & (kms$^{-2}$kpc$^{-1}$)                  &       &                           \\
\hline
IC5249 &1969.26$\pm$454.57&2.15$\pm$0.58&5.68 \\
UGC7321&7551.75$\pm$937.48&1.39$\pm$0.26&1.92\\
IC2233&13951.59$\pm$1244.02&0.11$\pm$0.01&$\sim$1239\\	
\hline
\end{tabular}}
\hfill{}
\label{tb:tablename}
\end{minipage}
\end{center}
\end{table*}

\noindent Interestingly, all the mass models with an NFW DM halo give poor fits with unphysical values for the best-fitting parameters.
This may possibly be understood on the basis of the fact that the cuspy
NFW profile fails to fit slowly-rising, dwarf-like rotation curves in general 
(See, for instance, Oh et al. 2008).
In Figure 7, we plot the observed rotation velocities of our sample galaxies plus those of IC2574 and IC2366 (Oh et al. 2008) as a function of 
the galacto-centric radius $R$
normalised with respect to the exponential stellar disc scale-length $R_D$ ($R_D$(1) for our sample galaxies). It clearly shows that the 
rotation curves of IC2233, IC2574 and
IC2366 rise much more slowly with respect to the stellar disc i.e., the rotation curve continues to rise linearly beyond one $R_D$, in contrast 
with those of the IC5249 and UGC7321, for which the linear region of the rotation curve is confined to $R$ $<$ $R_D$. \\ 

\begin{figure*}
\begin{center}
\begin{tabular}{c}
\resizebox{50mm}{!}{\includegraphics{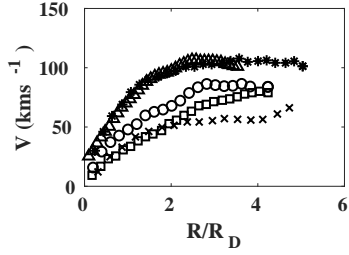}}\\
\end{tabular}
\end{center}
\caption{Comparison of the observed rotation curves of our sample galaxies 
IC5249(\emph{triangles}), UGC7321(\emph{stars}),IC2233(\emph{open circles})) plus those of IC2574 (\emph{crosses}) and IC2366 (\emph{squares}) 
(Oh et al. 2008) plotted as a function of the galacto-centric radius $R$ normalized with respect to the radial scale-length of the outer disc 
$R_{D}(2)$}.
\label{fig:slope}
\end{figure*}

\noindent The results of our best-fitting mass models with PIS and NFW DM halo are summarised in Tables 5 and 6 respectively. \\ \\

\noindent \textbf{Comparison with the dark matter halo profiles of LSBs:} In Figures 8 and 9, we compare our best-fitting model parameters with  
with PIS and NFW DM halo density profiles for our sample superthins with those for a sample of low surface brightness galaxies with 
similar dynamical mass and surface brightness from de Blok et al. (2001) for different cases: Constant ${\Upsilon}*$ case, Maximum Disc Case 
(only for PIS DM haloes) and Minimum Disc Case.
For the PIS DM halo case (Figure 8), we find that our best-fitting values lie well within the range values
 obtained for LSB galaxies except for the Maximum Disc case, for which our $R_c$ values are found to lie beyond the range of 
 values predicted for LSBs. However, the current work has shown that the Maximum Disc case in general gives poor fits to the observed rotation 
curves of superthins, and therefore may not be quite representative of the actual underlying profile of their DM haloes.
For the NFW case (Figure 9), DM halo parameters of our sample superthins compare well with those of the LSBs for the Constant ${\Upsilon}*$ and 
the Minimum Disc cases. Besides, we do not show the Maximum disc case here as the models were ruled out being physically not meaningful. 
For similar reasons, the parameters for IC2233 were not included in the plots.

\subsection{MOND} In Figure 10, we compare the best-fitting rotation curve using MOND with the observed rotation curves of IC5249,
UGC7321 and IC2233. With MOND, we obtain a reasonable best-fitting model to the rotation curve both for
IC5249 and UGC7321 with acceptable values of A and  ${\Upsilon}*$, the latter being close to the values dictated by the population synthesis 
models. For IC2233, however, we obtain physically unrealistic values of the best-fitting parameters A and ${\Upsilon}*$. 
However, one may also fits to the observed rotation curve by letting the distance to the galaxy $D$ and its inclination $i$ as free parameters (McGaugh \& de Blok 1998, Gentile et al. 2013, Angus et al. 2015). We defer this analysis to a
future paper. The results of our best-fitting mass models with MOND are summarised in Tables 7.

\begin{figure*}
\begin{center}
\begin{tabular}{ccc}
\resizebox{50mm}{!}{\includegraphics{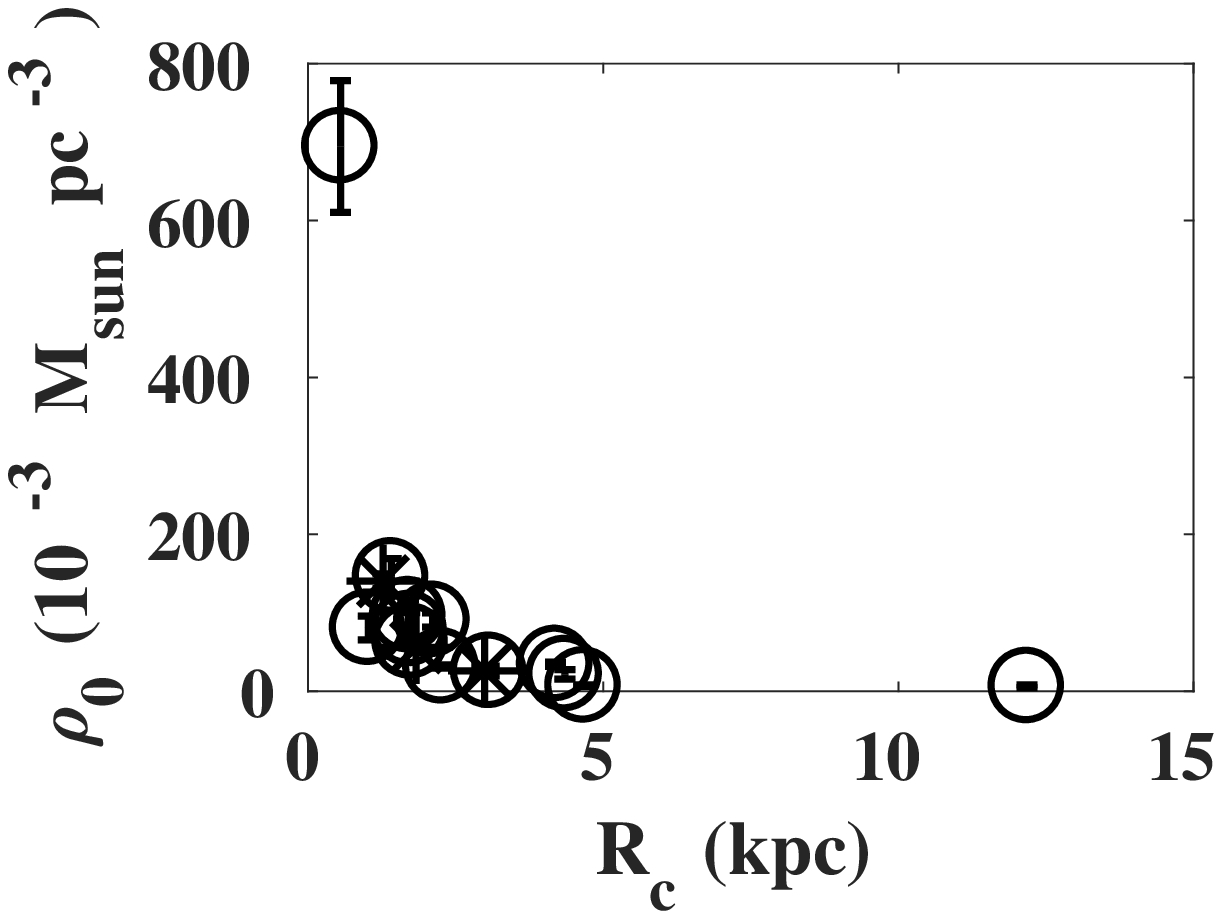}} &
\resizebox{50mm}{!}{\includegraphics{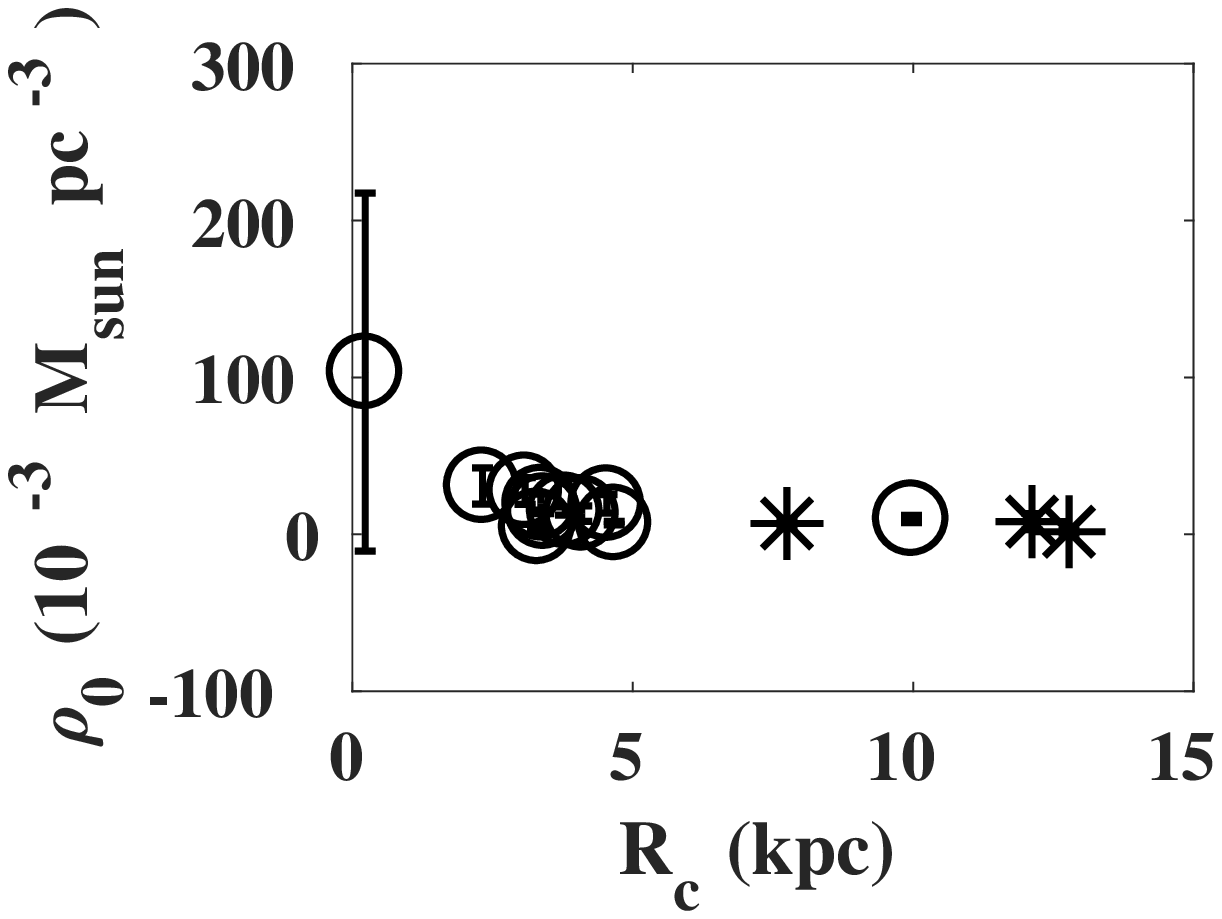}} &
\resizebox{50mm}{!}{\includegraphics{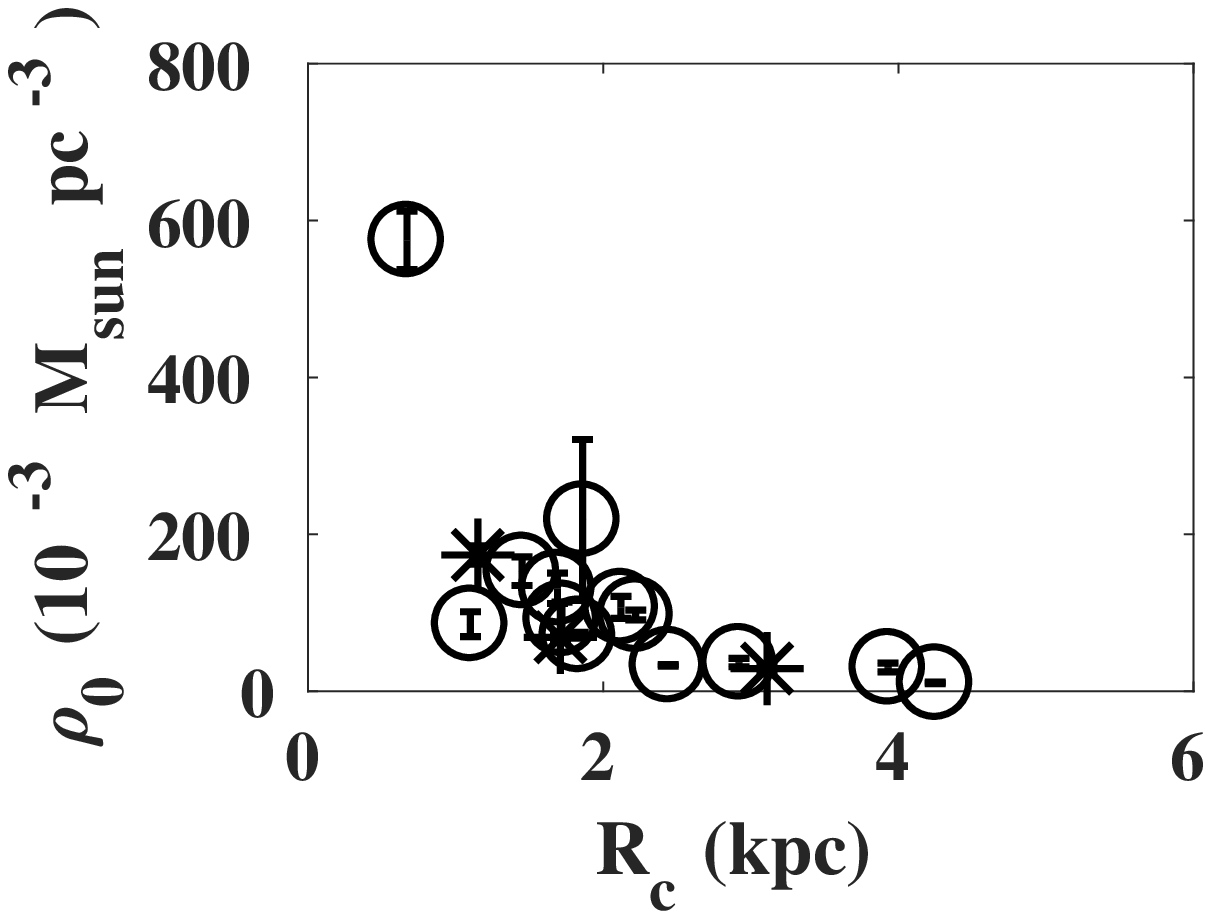}} \\
\end{tabular}
\end{center}
\caption{Comparing the best-fitting values of ${\rho}_0$ and $R_c$ of the PIS DM halo model for our sample of superthin galaxies (\emph{crosses}) with
those of a sample of low surface brightness galaxies with similar dynamical mass and surface brightness (de Blok et al. 2001) 
(\emph{open circles}):
Panel [1] Constant ${\Upsilon}*$ case as predicted by stellar population synthesis studies, Panel [2] Maximum disc case and Panel  
[3] Mimimum disc case}
\label{fig:lsb_pis}
\end{figure*}

\begin{figure*}
\begin{center}
\begin{tabular}{ccc}
\resizebox{50mm}{!}{\includegraphics{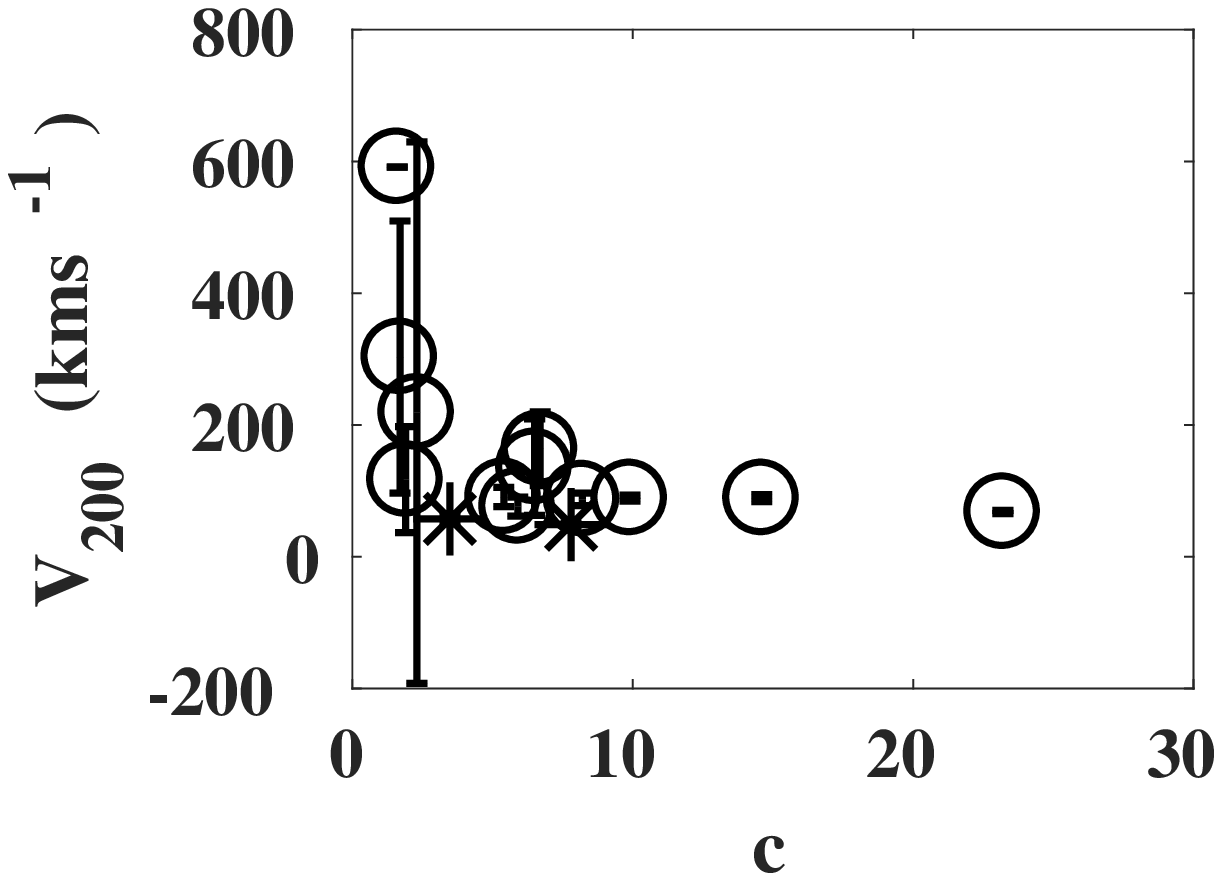}} &
\resizebox{50mm}{!}{\includegraphics{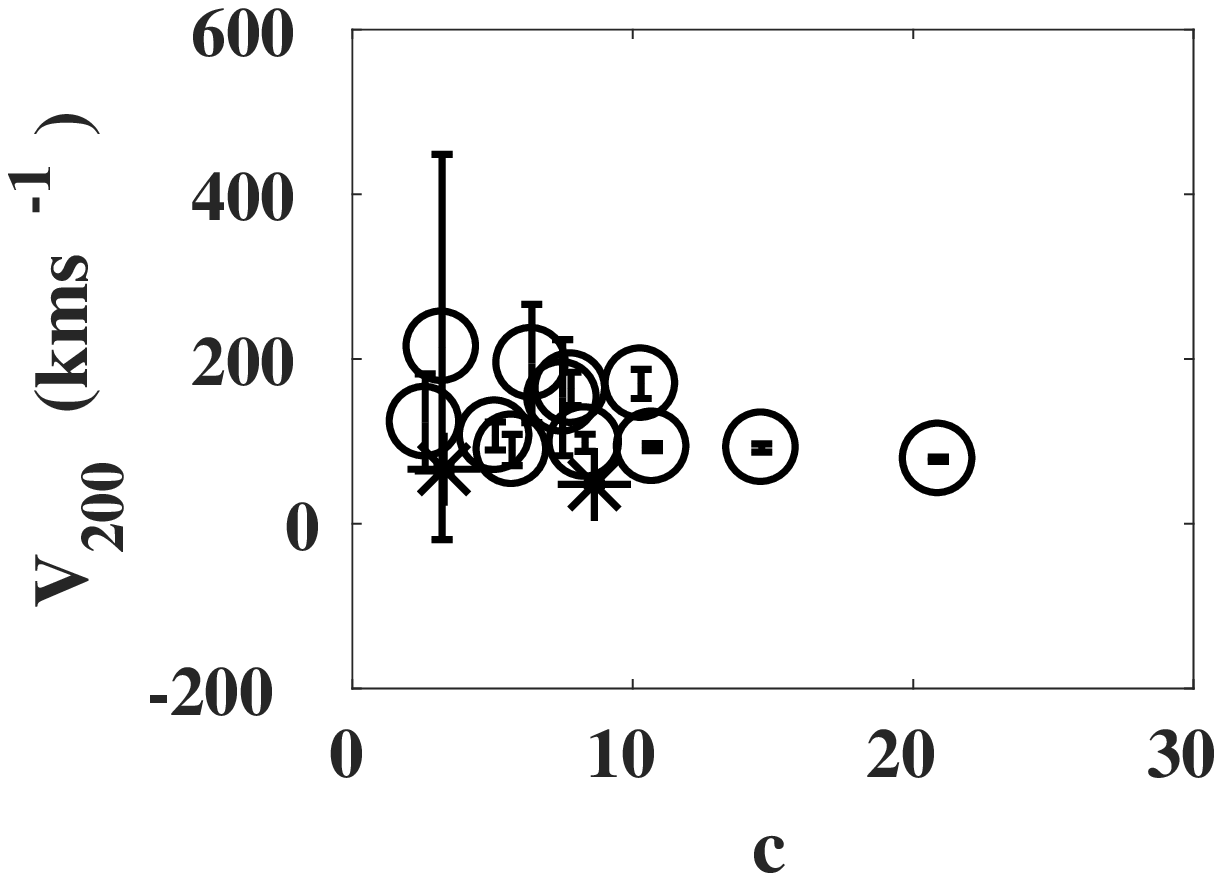}} \\
\end{tabular}
\end{center}
\caption{Comparing the best-fitting values of $c$ and $V_{200}$ of the NFW DM halo model for our sample of superthin galaxies  
(\emph{crosses}) with those of a sample of low surface brightness galaxies with similar dynamical mass and surface brightness 
(de Blok et al. 2001) (\emph{open circles}) : Panel [1] Constant ${\Upsilon}*$ case as predicted by stellar population synthesis studies, Panel [2] Minimum disc case}
\label{fig:lsb_nfw}
\end{figure*}

\begin{figure*}
\begin{center}
\begin{tabular}{ccc}
\resizebox{40mm}{!}{\includegraphics{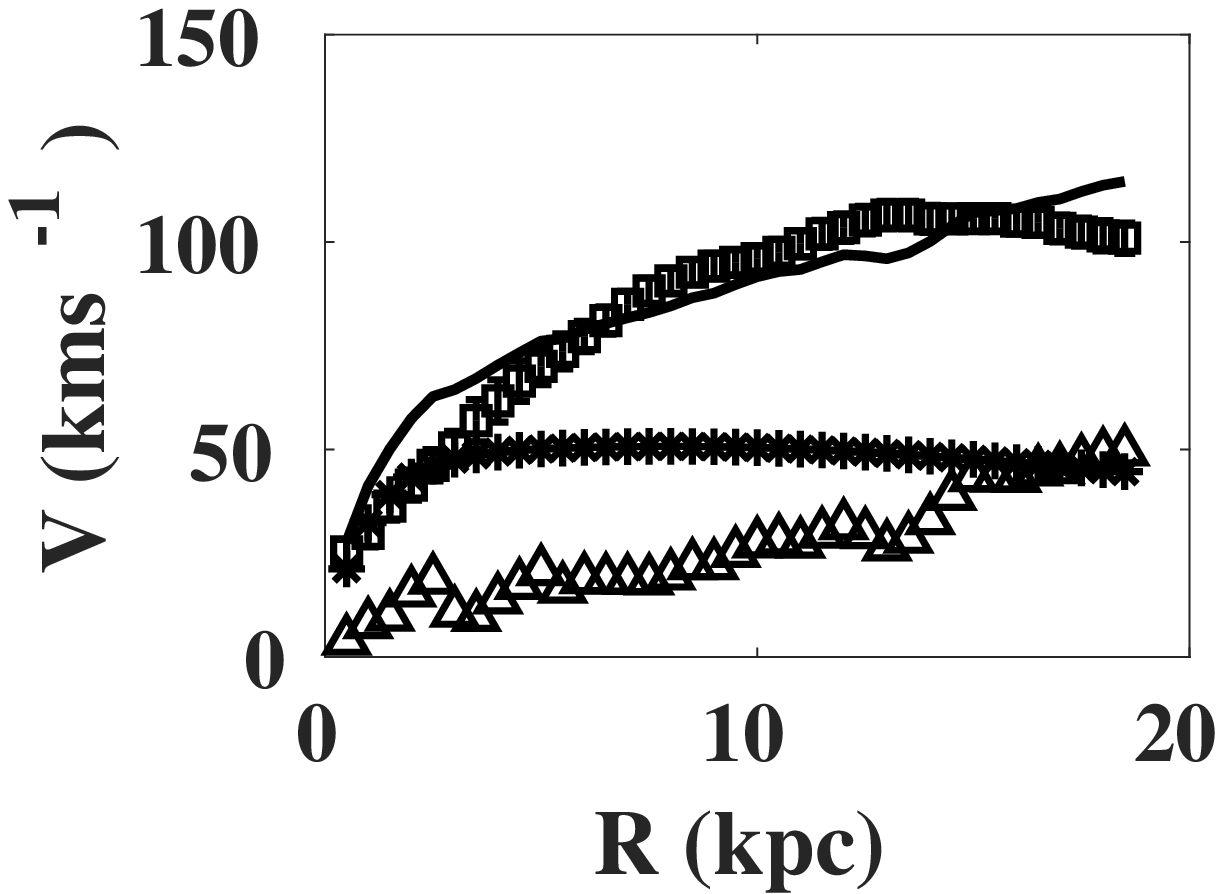}} &
\resizebox{40mm}{!}{\includegraphics{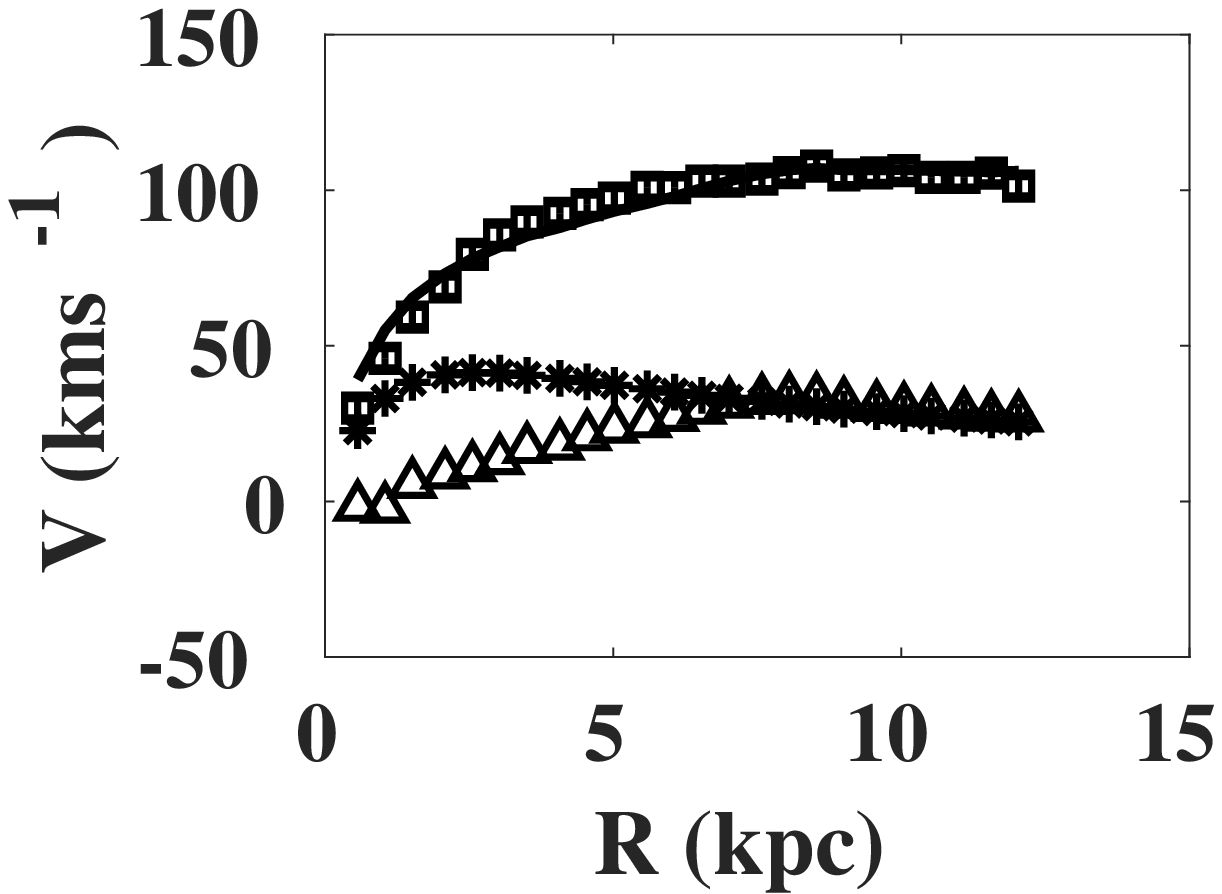}} &
\resizebox{40mm}{!}{\includegraphics{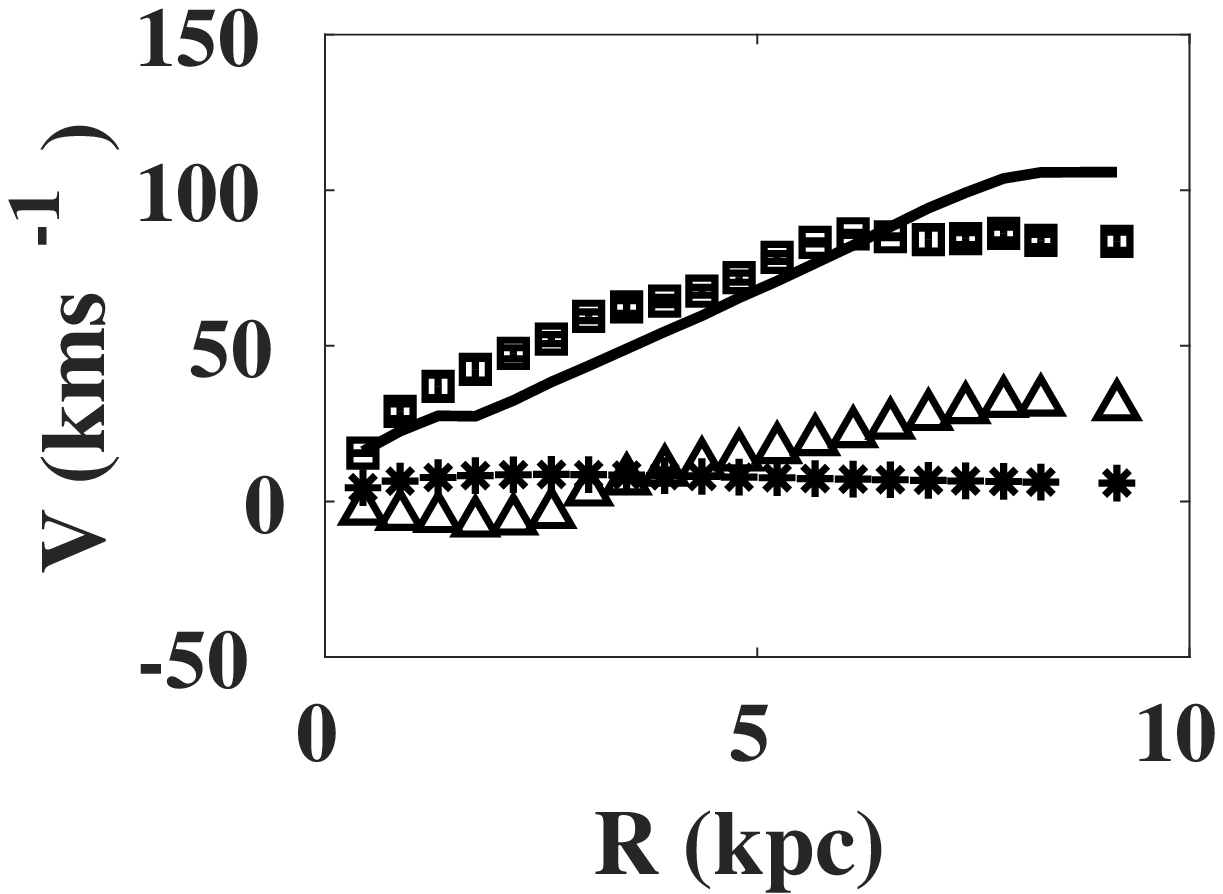}} \\
\end{tabular}
\end{center}
\caption{Modelling the HI Rotation Curve of the superthin galaxies: Panel [1] IC5249 Panel [2] UGC7321 and Panel [3] IC2233 
using MOND.
The \emph{stars} indicate the rotation curve due to the stellar disc alone, the \emph{triangles} the rotation curve due
to the gas disc, the \emph{solid line} the best-fitting model rotation curve and
the \emph{squares} the observed rotation curve with error-bars.}
\label{fig:mond}
\end{figure*}
\section{Conclusions}

\noindent We construct mass models of three superthin galaxies, U7321, IC5249 and IC2233, using high resolution rotation curves and gas surface 
density distributions obtained from HI 21 cm radio-synthesis observations, in combination with their two-dimensional structural surface brightness decompositions at Spitzer 3.6 $\mu$m band, all of which were available in the literature. We find that both
pseudo-isothermal and NFW-like dark matter density profiles give equally good fits to the observed rotation curves of IC5249 and UGC7321 but
the NFW profile fails to comply with the slowly rising rotation curve of IC2233; while the linear 
 regime of the observed rotation curve is confined within two radial disc scale-lengths ($R < 2 R_D$) for IC5249 and UGC7321, it 
extends beyond $R > 2 R_D$ for IC2233. Similarly, mass models with MOND give good fits to the rotation curves of IC5249 and UGC7321 but not 
with that of IC2233. Interestingly, however, for all of our sample galaxies, the best-fitting mass model with a PIS dark matter
density profile indicate a {\it compact} dark matter halo i.e., $R_c/R_D$ $<$ 2 where $R_c$ is the core radius of the dark matter halo
 which, in turn, may be fundamentally responsible for the existence of superthin stellar 
discs in these LSB galaxies. Our results may have important implications for the 
formation and evolution of models of superthin galaxies in the universe.

\clearpage

\section*{References}
Angus, G. W. et al., 2015, MNRAS, 451, 3551 \\
Banerjee, A. et al., 2010, NA, 15, 89 \\
Banerjee, A., Jog, C. J., 2011, ApJ, 732, L8 \\
Banerjee, A., Jog, C. J., 2013, MNRAS, 431, 582 \\
Bell, E. F., de Jong, R. S., 2001, 550, 212 \\
Binney, J., Tremaine, S., 1987, Galactic Dynamics. Princeton Univ. Press, Princeton, NJ \\
Bizyaev, D. V. et al., 2014, ApJ, 787, 24 \\
Byun, Y.-I., 1998, ChJPh, 36, 677 \\
Casertano, S., 1983, MNRAS, 203, 735 \\
Comeron, S. et al., 2011, ApJ, 741, 28 \\
Dalcanton, J. J., Bernstein, R. A., 2002, AJ, 124, 1328 \\
de Blok, W. J. G. et al., 2001, AJ, 22, 2396 \\
de Blok, W. J. G. et al., 2008, AJ, 136, 2648 \\
Gentile, G. et al., 2013, A\&A, 554, 125 \\
Goad, J.W., Roberts, M.S., 1981, ApJ, 250, 79 \\
Karachentsev, I. D. et al., 1993, AN, 314, 97\\
Karachentsev, I. D. et al., 1999, BSAO, 47, 5 \\
Kautsch, S. J., et al., 2006, A\&A, 445, 765\\
Kautsch, S. J.,  2009, PASP, 121, 1297 \\
Kregel, M. et al., 2005, MNRAS, 358, 503\\
Matthews, L. D. et al., 1999,  ASPC, 182, 223 \\
Matthews, L. D. et al., 1999, AJ, 118, 2751 \\
Matthews, L. D., 2000, AJ, 120, 1764 \\
Matthews, L. D., van Driel, W., 2000, A\&AS, 143, 421 \\
Matthews, L. D., Wood, K. 2003, ApJ, 593, 721 \\
Matthews, L. D. et al., 2005, AJ, 129, 1849 \\
Matthews, L. D., Uson, J. M., 2008, AJ, 135, 291 \\
McGaugh, Stacy S., de Blok, W. J. G., 1998, ApJ, 499, 66 \\
Milgrom, M., 1983, ApJ, 270, 365 \\
Navarro, J. F. et al., 1996, ApJ, 462, 563 \\
O'Brien, J. C. et al., 2010a, A\&A, 515, 60 \\
O'Brien, J. C. et al., 2010b, A\&A, 515, 61 \\
O'Brien, J. C. et al., 2010c, A\&A, 515, 62 \\
O'Brien, J. C. et al., 2010d, A\&A, 515, 63 \\
Oh, S.-H. et al., 2008, AJ, 136, 2761\\
Oh, S.-H. et al., 2015, AJ, 149, 180 \\
Sackett, P. D., 1997, ApJ, 483, 103\\
Salo H., et al., 2015, ApJS, 219, 4 \\
Uson, J. M. \& Matthews, L. D., 2003, AJ, 125, 2455 \\
van der Kruit, P. C. et al., 2001, A\&A, 379, 374 \\
Yoachim, P., Dalcanton, J. J., 2006, AJ, 131, 226\\

\section*{Acknowledgements}

\noindent AB would like to thank Kaustubh Waghmare and Jayaram Chengalur for useful discussion and suggestions,
 Erwin de Blok and Hans Terlouw for help with GIPSY and finally the anonymous referee whose critical comments have greatly helped
 to improve the quality of the paper.

\end{document}